\RequirePackage{fix-cm}
\documentclass[twocolumn]{svjour3}          
\smartqed  
\usepackage{graphicx}
\usepackage{epstopdf}
\usepackage{natbib}
\bibpunct{(}{)}{;}{a}{}{;}
\usepackage[utf8]{inputenc} 
\usepackage{graphicx}
\usepackage{color}
\usepackage{float}
\usepackage{xcolor}
\usepackage{ulem}
\usepackage{amsmath}
\usepackage{subfig}
\usepackage{mathtools}
\usepackage{amssymb}
\usepackage{natbib}
\usepackage{mathrsfs}
\usepackage{commath}
\usepackage{mathtools}

\let\*=\times

\usepackage{ulem}
 
\begin{document}

\title{Variational Mode Decomposition for estimating critical reflected internal wave in stratified fluid}



\author{Horne E.$^{1,2,\text{\textdagger}}$ \and Schmitt J.$^{2}$ \and Pustelnik N.$^{2}$ \and Joubaud S.$^{2,3}$ \and Odier P.$^{2}$}

\institute{
$^{1}$ LadHyX, CNRS, \'Ecole Polytechnique, 91128 Palaiseau CEDEX, France.\\
$^{2}$ Univ Lyon, ENS de Lyon, Univ Claude Bernard, CNRS, Laboratoire de Physique, F-69342 Lyon, France. \\
$^{3}$ Institut Universitaire de France (IUF), F-75005 Paris, France. \\
$^\text{\textdagger}$ \email{ernesto.horne@ladhyx.polytechnique.fr}}

\date{\today~Received: date / Accepted: date}

\maketitle

\begin{abstract}

The shear resulting from internal wave reflections can play a crucial role in the transport and resuspension of sediments in oceanic conditions. In particular, when these waves undergo a \textit{critical reflection} phenomenon, the reflected wave can produce a very large shear. Separating the reflected wave from the incident wave is a technical challenge since the two waves share the same temporal frequency. In our study, we present a series of experimental measurements of internal waves in \textit{critical reflection} configuration and we analyze them using the 2D-VMD-prox decomposition method. This decomposition method was adapted to specifically decompose waves in an internal wave critical reflection, showing an improvement in its performance with respect to preexisting internal wave decomposition methods. Being able to confidently isolate the reflected wave allowed us to compare our results to a viscous and non-linear model for critical reflection, that correctly describes the dependence of the shear rate produced in the boundary as a function of the experimental parameters.

\keywords{Internal waves critical reflection \and  sediment resuspension \and  wave decomposition}
\end{abstract}

\section{\label{sec:reflection:introduction}Introduction}

\begin{figure}[h]
\centering
\includegraphics[width=7cm]{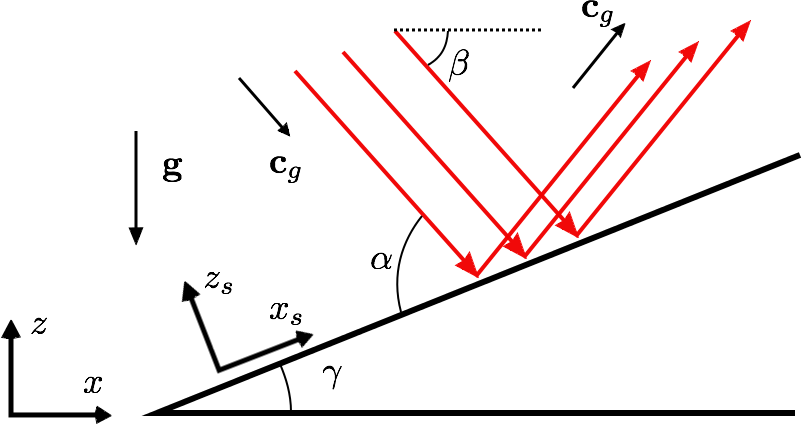} 
\caption{Schematic view of an internal wave reflection. The angle between the bottom slope and the horizontal is $\gamma$; the angle between the incident group velocity and the horizontal is $\beta$, and $\alpha=\gamma+\beta$. ${\mathbf c}_g$ indicates the group velocity and ${\mathbf g}$ is the acceleration of gravity. The horizontal and vertical coordinates $(x,z)$, as well as the coordinates attached to the slope, $(x_{s},z_{s})$, are indicated.}
\label{figure1}
\end{figure}

Internal gravity waves are omnipresent in stratified fluids such as seas and oceans, atmosphere or planetary interiors. The primary mechanism leading to the generation of internal waves in the ocean interior is the interaction of global tides with the bottom topography. In such stratified fluids with an initially constant buoyancy frequency $N =\left[-g\partial_{z}\rho_0(z)/\bar{\rho}\right]^{1/2}$, where $g$ is the acceleration of gravity, $\rho_0(z)$ is the density of the fluid at rest and $\bar{\rho}$ a reference background density, internal waves propagate obliquely at an angle $\beta$ with respect to the horizontal according to the dispersion relation
 \begin{equation}
 \omega = N \sin\beta\,
 \end{equation}
where $\omega$ is the forcing frequency. This peculiar dispersion relation requires the preservation of the angle $\beta$ upon reflection on a rigid boundary. In the case of a  boundary tilted at an angle $\gamma$ with respect to the horizontal, this purely geometric property can lead to strong variations of the width and amplitude of the wave (focusing or defocusing) upon reflection as illustrated in Figure~\ref{figure1}~\citep{Phillips:66}.

Internal-wave focusing leads to large shear, bottom layer instabilities and in some cases overturning \citep{Buhler2007,Zhang:PRL:08,Gayen2010,Chalamalla2013,SarkarARFM2017}. In addition, a transfer of energy through scales develops as a consequence of non-linearities  \citep{Brouzet17,Dauxois17} producing vertical mixing and mean flows. From a geophysical point of view, a precise description of internal wave reflections is key for understanding the vertical mixing in the ocean~\citep{Ivey:ARFM:2008} and its effect as a mechanism for sediment transport and resuspension~\citep{Cacchione2002}. The latter has been reported in many observational studies~\citep{Bogucki:JPO:97, Quaresma:MG:07, Hosegood:GRL:04,Butman:CSR:06}, showing that internal gravity waves are a first order mechanism for sediment resuspension.

Of special interest is the case of the \textit{critical reflection}, occurring when the angle of the slope is equal to the angle of propagation of the internal wave, \textit{i.e.} $\beta=\gamma$. This particular reflection occurs as a transition between the state for which the reflected wave propagates up the slope ($\beta > \gamma$) and the state for which the wave propagates down the slope ($\beta < \gamma$). 
A linear theory developed by~\cite{Phillips:66} was the first intent to describe internal wave reflections, nevertheless, it predicts a divergence of the energy of the reflected wave when reaching critical angle ($\gamma=\beta)$. This is contrasted with what has been recognized in experiments and observations~\citep{Cacchione:JFM:74,DeSilva:JFM:97,Cacchione2002,Gostiaux:PoF:06}, indicating the existence of a mechanism that prevents the singularity from developing. 
A balance between non-linearities and dissipation appears as the most complete description for the underlying mechanism preventing the singularity. Several theoretical models have been proposed. \citet{Wunsch:JFM:69} added a friction term to the linear inviscid solution allowing the singularity to heal thanks to viscous effects in a boundary layer in the surrounding of the slope. \citet{Thorpe:JFM:1987} suggested that the spatial overlap between the linear inviscid solution of the incident and reflected waves generates resonance of higher harmonics that heal the singularity. \citet{Kistovich:JAMM:95} proposed that, in the linearized equations, viscosity and diffusion restrict the limiting value of the geometrical compression coefficient of the reflected beam. \citet{Dauxois:JFM:99} developed a solution taking into account non-linearities and viscosity for the critical reflection, while developing an inviscid solution for near-critical reflection. They where able to heal the singularity by using a temporal description of the wave field and, thanks to a matched asymptotic expansion, to solve the non linear equation.  \citet{Scotti:JFM:11} also computed a solution for which the non-linearities heal the singularity of Phillips linear solution. Nevertheless, his solution is presented only for inviscid fluids whereas, in contrast to oceanic conditions, viscosity plays an important role for internal waves studied in the laboratory. In this work our results are compared with the solution of \citet{Dauxois:JFM:99}. This theoretical model presents the most complete solution for the critical reflection, as viscous dissipation and non-linearities are included.

With the development of experimental techniques to generate well defined internal waves and of observation methods in the laboratory, some of these theoretical predictions have been confronted with experimental data. \citet{Dauxois:PoF:04} measured the density profile close to the slope using synthetic Schlieren technique finding a qualitative agreement with the theoretical predictions of~\citet{Dauxois:JFM:99}. With the same measurement technique, \citet{Peacock:PoF:05} evidenced second harmonic generation in the reflection, due to non-linear processes and \citet{Gostiaux:PoF:06} confirmed this observation with quantitative velocity measurements using PIV. The wavelength selection taking place in the reflection process is analyzed in~\citet{Tabaei:JFM:05} and in~\citet{Dauxois17} showing that the spatial overlap between the incident and the reflected wave and the geometrical configuration of the reflection process strongly influence the waves produced through non-linearities. \citet{Zhang:PRL:08} have experimentally studied wave generation over an oscillating topography observing that the waves are generated in a near critical region. These waves produce very strong shear near the boundary and therefore can become unstable and break. Using two fitting parameters, they have shown a good agreement between their measurements and the viscous solution of \citet{Dauxois:JFM:99}.

To go beyond the previous results and investigate wave reflection at criticality, it is important to characterize independently the reflected wave as well as the incident wave. However, close to criticality, the separation through Hilbert transform, as proposed in \citet{Mercier:PoF:08}, struggles to fully separate the incident and the reflected wave as a consequence of the overlap that these two waves present in their spatial spectrum. Several methods have been recently introduced to properly deal with mode decomposition going from \textit{Empirical Mode Decomposition} (EMD) \citep{Huang:1998,Rilling:2003} to synchrosqueezing \citep{Daubechies:2011}. If the first class of approach performs well with very few prior information, except the number of modes, it lacks a theoretical point of view and has no convergence guarantee. Thus, the decomposition steps are very sensitive to noise and to sampling~\citep{Pustelnik:2014}. On the other hand, synchrosqueezing has a strong theoretical framework based on wavelets but it requires strong prior on the location of the modes. A good compromise between the two approaches, named \textit{Variational Mode Decomposition} (VMD), is developed in \citet{Dragomiretskiy2014,Zosso2017}. The objective of this work is to improve over this recent and efficient mode decomposition in order to deal with the PIV data for internal waves measured near a critical or close to critical slope.

\begin{figure*}[h]
\centering
\includegraphics[width=11cm]{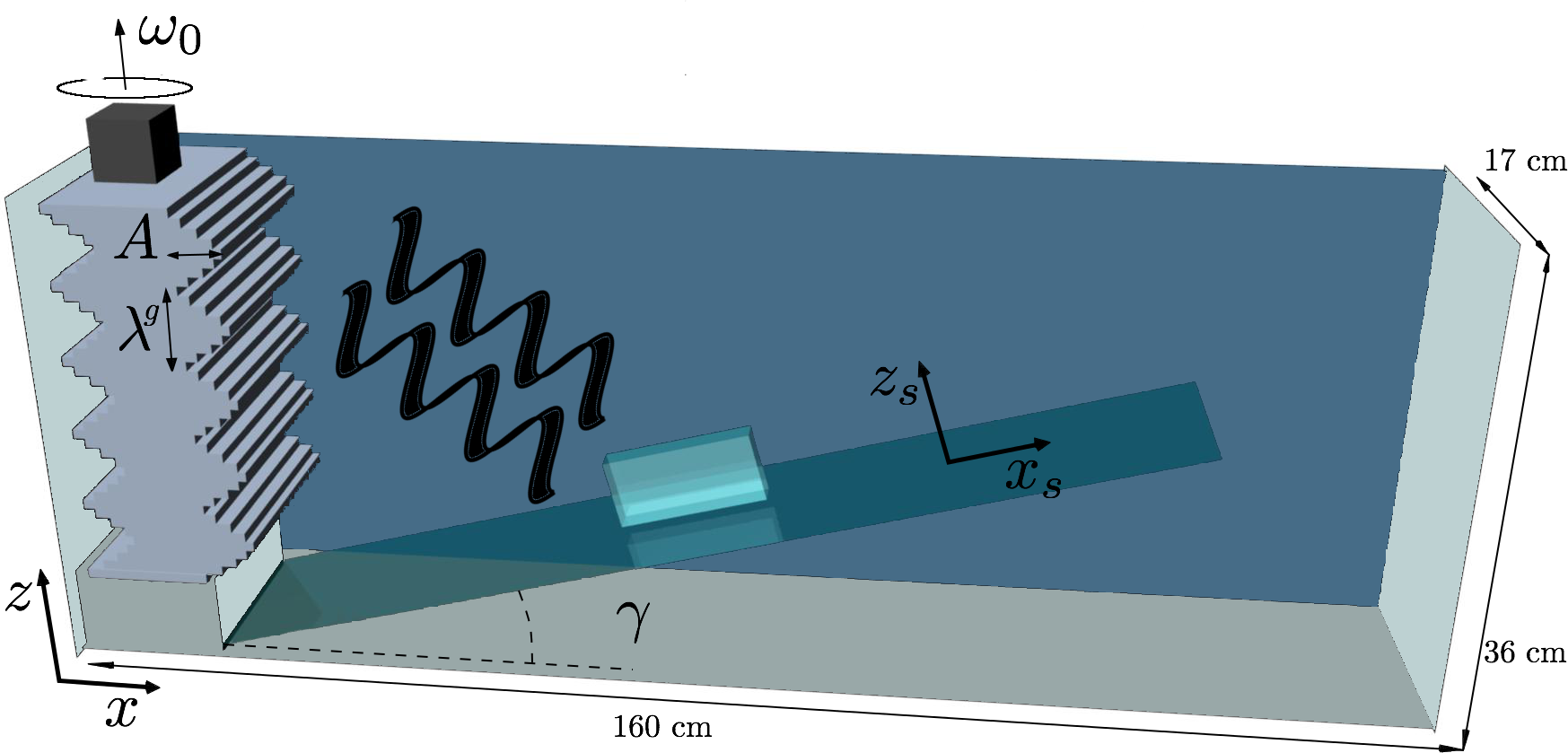} 
\includegraphics[width=3cm]{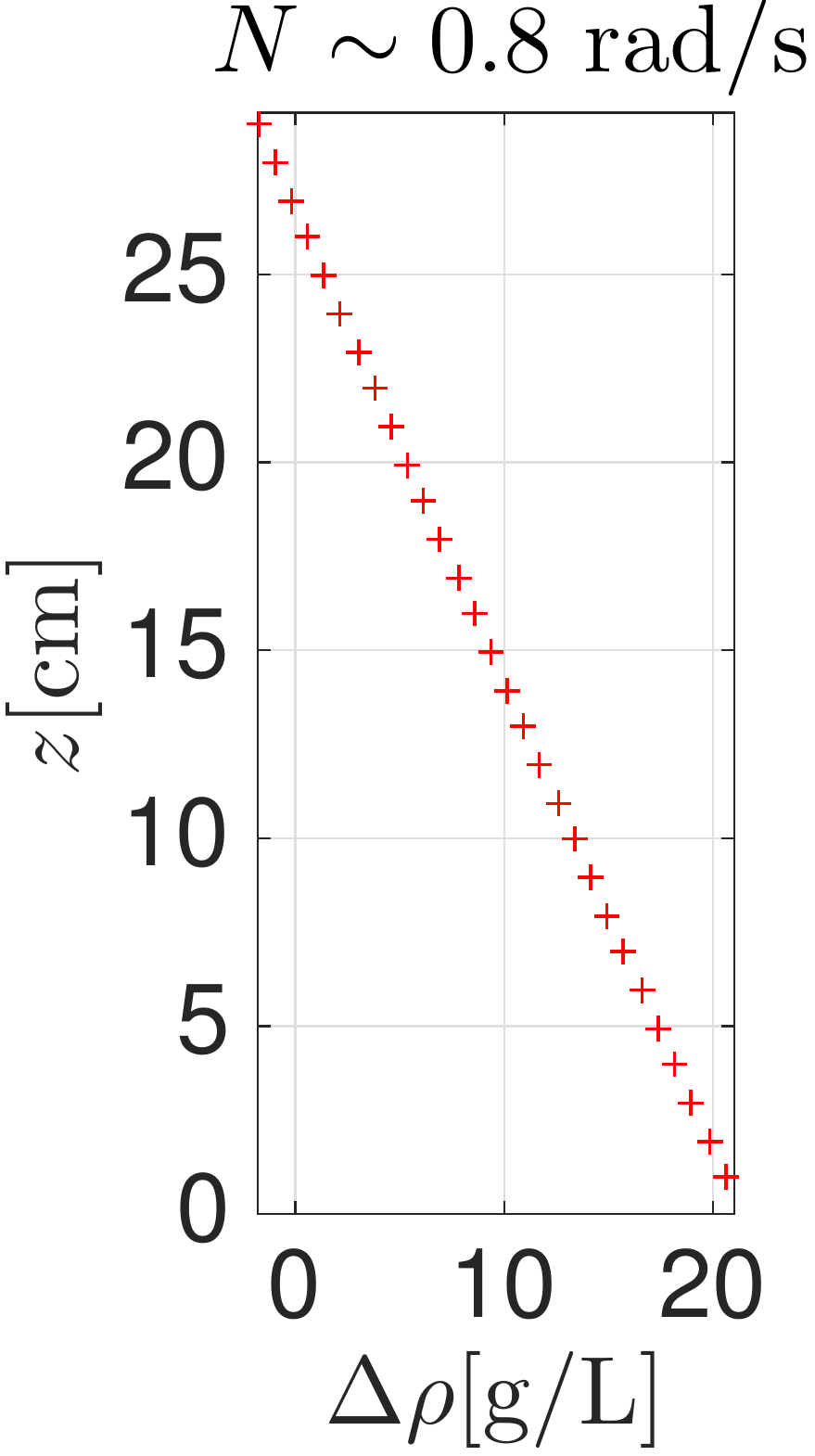} 
\caption{(Left) Sketch of the experimental setup. The plane-wave generator is on the left of the tank. The internal waves propagate from up-left to down-right and reflect on the inclined slope. The field of view is represented with a rectangle parallel to the slope. The control parameters are indicated in the sketch: $A$, $\lambda^{\mathrm{g}}$, $\omega_0$ and $\gamma$. (Right) Experimental measurement of the density as a function of height.}
\label{figure2}
\end{figure*}

In the next section, we present the experimental set-up and describe the classical data processing method used to extract the reflected wave from the measured velocity field, commenting on its limitations near criticality. VMD and the proposed adaptations to study critically reflected internal waves are presented in \S~\ref{sec:VMD}. In \S~\ref{sec:appli} the application of the proposed 2D-VMD-prox method over a synthetic critical reflection is used to optimize the parameters and highlight the benefits and limitations of the method. The results of the decomposition over our internal wave critical reflection data and their comparison with the theoretical predictions are then exposed. Conclusions are drawn in the last section.
 
\section{\label{sec:reflection:system}Experimental setup} 

\subsection{General setting}

Experiments are performed in a tank 160~cm long, 17~cm wide and 42~cm deep, filled with 36~cm of salt water as sketched in Figure~\ref{figure2}(Left). Using the two-bucket method~\citep{Fortuin:JPS:60,Oster:CR:63}, the fluid is linearly stratified in density in order to produce a constant buoyancy frequency $N$. Density vertical profile measurements are performed with a conductivity probe. An example of experimental measurement of density as a function of height is plotted on the right side of Figure~\ref{figure2}. A transparent acrylic slope with variable inclination is inserted in the tank before the filling procedure. The presence of a sloped boundary in a stratified fluid, associated with the no flux contraint for salt out of the boundary implies a local curvature of the isopycnals at this boundary. The thickness of this boundary region is of the order of $5 \cdot 10^{-2}$~cm \citep{Phillips:DSR:70}, which is beneath the experimental spatial resolution of the velocity field and we therefore neglect this effect in this study. The angle of the slope $\gamma$ can vary between $0^{\circ}$ and $35^{\circ}$, and the plate is 16 cm wide. The velocity of the fluid is measured using particle image velocimetry (PIV). A vertical laser sheet is produced by combining a laser and a rapidly oscillating mirror. Hollow glass spheres (10~$\mu$m diameter and 1.1 g $\cdot$ cm$^{-3}$ density) used as passive tracers are seeded in the fluid. Images of 2452 $\times$ 1452 pixels are taken representing a real size of 11.3 $\times$ 6.7 cm$^2$ for the experiment with large spatial resolution. The image processing analysis is performed by comparing two successive images, with a temporal resolution of 0.25~s. Each image is divided in boxes of 25 $\times$ 25 pixels that browse in an 80 $\times$ 80 pixel size box researching the maximum cross correlation for two successive images, which is used to compute the velocity vector in the $xz$ plane. 

The wave generator \citep{Gostiaux:EF:07,Mercier:JFM:10}, consisting of stacked moving plates, is set vertically at the left side of the tank such that the plates move horizontally generating plane waves. The displacement profile producing plane waves is
\begin{equation}
X(z,t)=A \sin\left ( \omega_0 t -k^{\mathrm{g}}_{z} z \right ),
\end{equation} 
where $\omega_0$ is the forcing frequency, $k^{\mathrm{g}}_{z}$ the vertical wave\-number of the generator and $A$ its displacement amplitude. For the experiments presented in this work the amplitude of the plane wave generator varies between 0.25 and 1.5 cm, and the vertical wavelength $\lambda^{\mathrm{g}}=2\pi/k^{\mathrm{g}}_{z}$ between $\sim$4 and $\sim$8 cm. 

The configuration used to study the reflection process is schematized in Figure~\ref{figure1}. The incoming wave propagates from up-left to down-right at an angle $\beta$ with respect to the horizontal, imposed by the frequency of the wave generator $\omega_0$. It reflects on the oblique slope tilted at an angle $\gamma$ with respect to the horizontal and propagates away from the slope. Two coordinate systems are used, $x$ and $z$ are the coordinates respectively perpendicular and parallel to gravity, and $x_s$ and $z_s$ are the coordinates respectively along and normal to the slope. In the coordinates attached to the slope, the velocity field is $\mathbf{u} = (u_s,w_s)$ corresponding respectively to the velocity component along the slope and normal to the slope. 

Since theoretical results~\citep{Dauxois:JFM:99} propose an analytical expression of the along-slope component of the velocity field, in this study, we focus on this component: we aim to approximate the field $u_s$ at the position $(x_s,z_s)$ and time $t$ with a signal $u_M(x_s,z_s,t)$ expressed as a sum of $J$ modes. This model considers that, within experimental errors, the internal wave field in the tank can be written as:
\begin{equation}
u_M(x_s,z_s,t) =  \sum_{j=1}^J m_{j}(x_s,z_s,t)\,,
\label{eq:modedec}
\end{equation}
with the $J$ modes $m_{j}$ written as
\begin{eqnarray}
m_{j}(x_s,z_s,t)&=& a_j(x_s,z_s,t)\cos\left(k_{x_s,j}(x_s,z_s,t) x_s \right.\nonumber\\ &+&\left.k_{z_s,j}(x_s,z_s,t) z_s + \phi(t)\right)\,,
\end{eqnarray}
where $a_j$ 
models the amplitude changes, the mean value of $k_{x_i,j}$ is close to the experimental wavenumbers, and $\phi$ is a phase term. The model is valid near the generator where only the incident wave is present $J=1$, or near the slope with at least the incident and the reflected wave, $J=2$. In this study, we focus on the region near the slope and on the stationary regime (several wave periods after the moment in which the incident wave first reflects on the slope) in order to insure the presence of both waves in the field of view. We therefore assume that the observation is a sum of the incident and of the reflected wave. 
We study different experimental configurations, where the characteristics of the incident wave, intensity, wavelength and angle of propagation  are controlled by the experimental parameters $A$, $\lambda^{\mathrm{g}}$ and $\beta = \sin^{-1}(\omega_0/N)$. The characteristics of the reflected wave depend on the characteristics of the incident wave and on the control parameter $\beta-\gamma$ which is a measure of the departure from criticality ($\beta=\gamma$).
The parameters specific to each experiment are summarized in Table~\ref{table:reflection:system:exp}. 

\begin{table*}[t]
\centering
\begin{tabular}{cccccccc}
Case &$ N$[rad/s]&$\gamma [\,^{\circ}]$& $A$[cm]  &$\lambda^{\mathrm{g}} $[cm]&$\beta [\,^{\circ}]$ ($ \pm 1^{\circ}$)&  Re   & Fr  \\
\hline
exp1 & 1.15 & 16.5 & 0.25 & 4&12 - 25 & 6 &0.011\\
exp2 &1.14 &15 & 0.25& 8 &12 - 26& 31  &0.016\\ 
exp3 & 1.22 &15  & 1  &4 &8 - 22 & 21 &0.044\\ 
exp4  &1.1  & 16   & 1   &4  &7 - 33 & 22  &0.05 \\ 
exp5  &1.1 & 16    &1.5  &4 &7 - 34  &28 & 0.06\\
exp6 &1.13 & 16    &0.5 &8 &7 - 32   &46  &0.03\\ 
exp7 & 1.02 &17  &1  &8 &11 - 40  &106   &0.065\\

\hline 
\end{tabular}
\caption{\label{table:reflection:system:exp}Experimental control parameters. $N$ is 
the buoyancy frequency, $\gamma$ the angle of the slope, $A$ and $\lambda^{\mathrm{g}}=2\pi/k^{\mathrm{g}}_{z}$ are the amplitude and the vertical wavelength of the wave generator, $\beta$ indicates the explored range of angles of propagation of the incident wave given by the dispersion relation $\omega_0=N\sin \beta$, where $\omega_0$ is the forcing frequency of the generator. The \textit{critical reflection} corresponds to $\vert \beta-\gamma\vert=0 \pm 1^{\circ}$. $\textrm{Re}=U \lambda^{\mathrm{g}} /\nu$ is the Reynolds number, and $\textrm{Fr}=U/(\omega_0 \lambda^{\mathrm{g}} ) $ the Froude number, where $U$ is the maximum velocity in the direction of propagation of the incident wave. The Reynolds and Froude numbers are calculated for the cases of \textit{critical reflection}. 
}
\end{table*}

\begin{figure*}[t]
\centering
\includegraphics[width=11cm]{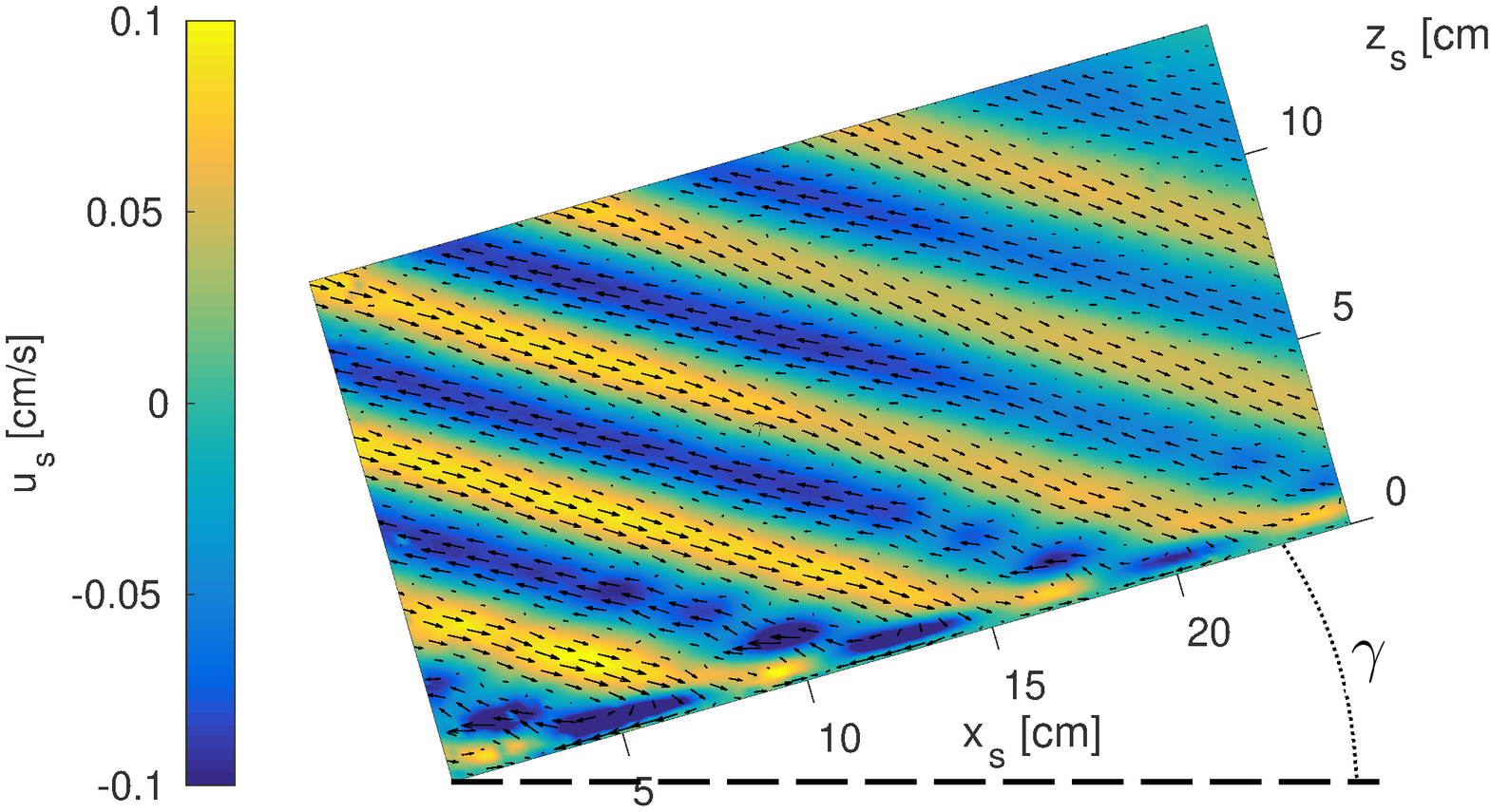}
\includegraphics[height = 5cm,width=6cm]{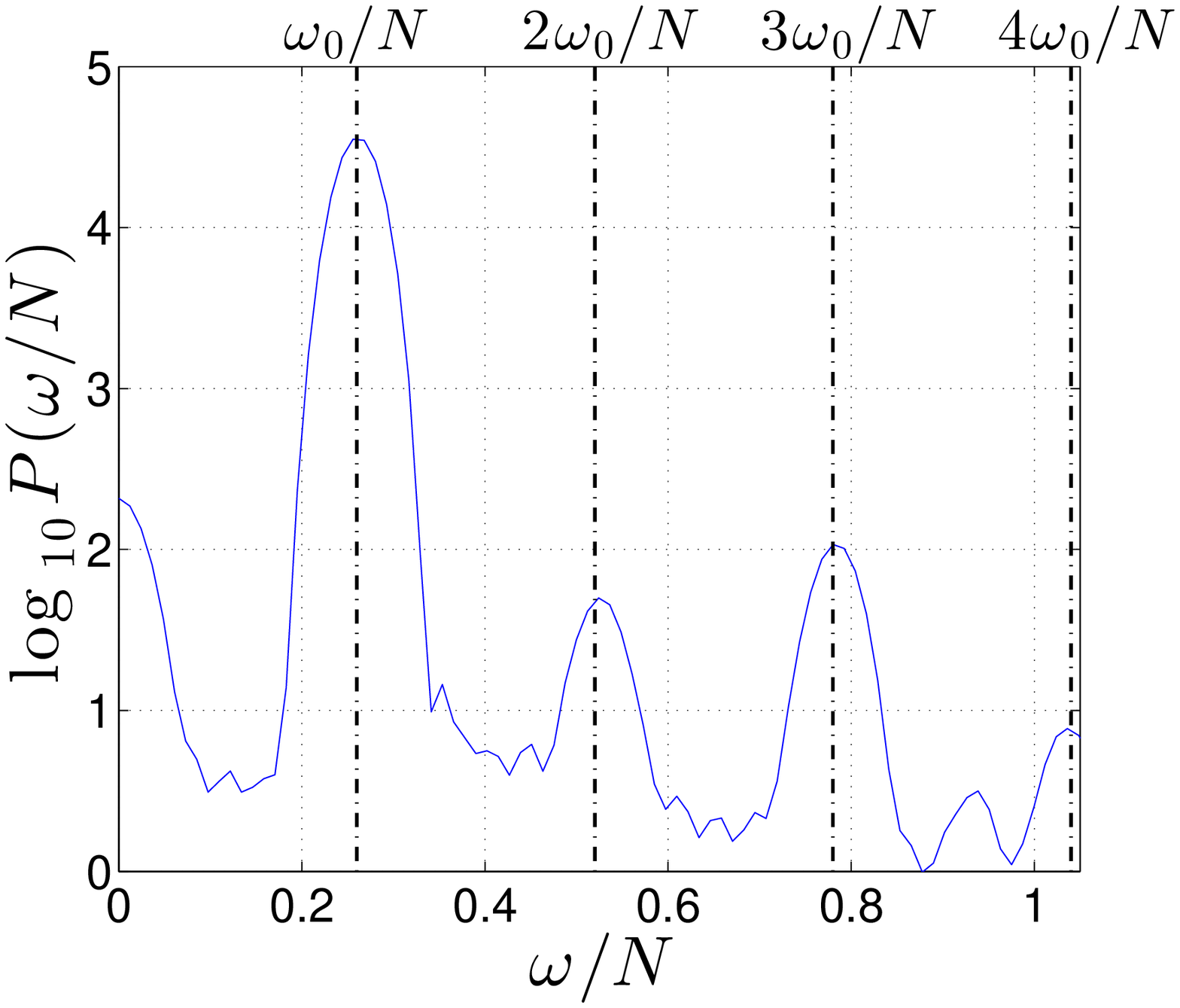} \\
\caption{(Left) Snapshot of the velocity field at $t/T_0 =15$ where $T_0 = 2 \pi/\omega_0$. The background color indicates the along-slope component of the velocity $u_s$, and the arrows represent the velocity field. 
$x_s$ and $z_s$ coordinates are indicated. The slope boundary is located at $z_s=0$. The incident wave is propagating from up-left to down-right, and the generator is located $\sim 30$ cm from the center of the image. The measurements correspond to ``exp3''. (Right) Associated power spectrum density $P(\omega/N)$. The spectrum is computed from a set of points located on the left side of the field of view at approximately 8 cm from the slope. The width of the peaks is related to the finite time span of the experiment, while the variations of the forcing frequency due to experimental limitations of the wave generator are much smaller.}
\label{figure3}
\end{figure*}

\subsection{Description of a typical experiment}

From the set of near-critical reflection experiments listed in Table~\ref{table:reflection:system:exp} we use ``exp3'' as an example to describe the critical reflection behaviour. The uncertainty on the departure from criticality is directly related to the uncertainty on $\abs{\beta-\gamma}$. The geometrical error related to $\gamma$ is small compared to the error on $\beta$ and therefore, is neglected. The error on $\beta = \sin^{-1}(\omega_0/N)$ is dominated by the intrinsic variations of the forcing $\omega_0$ and is of $\mathcal{O}(1^{\circ})$ for all experiments.   

A snapshot of the velocity field is shown in Figure~\ref{figure3}(Left), obtained $15$ periods after the wave generator was started. The incoming plane wave reflects on the sloped boundary (at $z_s=0$) and propagates along the slope in a narrow region producing an increase in the intensity of the velocity field, as a consequence of focusing. After a transient regime ($\sim 10$ periods), the reflection process reaches a stationary regime. One can also observe a decay of the amplitude of the incoming waves as they propagate away from their source as a consequence of viscous dissipation. 

\subsection{Temporal filtering}

To investigate the reflection mechanism, it is important to separate the different waves involved in the process. To do so, the first step is to perform a temporal band-pass filter centered on the forcing frequency of the incident wave. Figure~\ref{figure3}(Right) shows the frequency power spectrum for the experiment corresponding to the snapshot presented in Figure~\ref{figure3}(Left). The largest contribution is associated to the frequency $\omega_0 /N=0.26$, which comes from the incident and the reflected wave. As expected and already studied~\citep{Tabaei:JFM:05,Rodenborn:2011,Kataoka:2020}, one can observe higher order frequencies ($2\omega_0$, $3\omega_0$, etc.) with lower magnitude and a non vanishing contribution to the signal at $\omega/N=0$. The latter corresponds to a mean flow. All these frequencies are generated through nonlinear interactions between the incident and reflected waves. By applying a temporal band-pass filter centered on the frequency $\omega_0 /N=0.26\pm 0.01$, the components related to the higher harmonics or the mean flow are removed, and the velocity field obtained will be solely associated to the incident and the reflected wave with frequency $\omega_0$. In the remainder of the article, we therefore drop the temporal dependency of $d$ to indicate that a temporal filtering has been previously performed. A given temporal phase is selected for the velocity field filtered in time.

\subsection{Spatial filtering}

Although the incident and reflected wave oscillate in time at the same frequency, they do not propagate in the same direction and and they do not have the same wavenumber. To be more precise the absolute value of the along-slope component of the wavenumber is conserved for all $x_s$, but not the normal to the slope component of the wavenumber, i.e.,

{\small{$$\begin{cases}
\mid k_{x_s,j=\{{\mathrm{refl}}\}}\mid(x_s,z_s=0)&= \quad \mid k_{x_s,j=\{{\mathrm{inc}}\}}\mid(x_s,z_s=0)\\
\mid k_{z_s,j=\{{\mathrm{refl}}\}}\mid(x_s,z_s=0)&\neq\quad\mid k_{z_s,j=\{{\mathrm{inc}}\}}\mid(x_s,z_s=0).\end{cases}
$$}}
The separation between incident and reflected wave through the spatial spectrum is therefore based on the discrimination between $k_{z_s,j=\{{\mathrm{inc}}\}}$ and $k_{z_s,j=\{{\mathrm{refl}}\}}$.

A classical way to apply a spatial spectrum separation is to use the technique based on the Hilbert transform~\citep{Mercier:PoF:08}, commonly used in previous studies where several waves are entangled~\citep{Bourget:JFM:13}. However, in a near-critical case, since the reflected wave is only present in a narrow region very close to the inclined plane, the spectrum of $k_{z_{s}}$ is substantially wide, making the filtering operation delicate. All attempts at such filtering result in a pollution of the filtered reflected wave by a non negligible residue of the incident wave as illustrated in~\cite{Schmitt:EUSIPCO:15}. In order to limit these effects, we propose to analyse the data with more specific signal processing tools. 

\section{\label{sec:VMD}Variational Mode Decomposition method}

\subsection{Mathematical framework}

When dealing with the Hilbert transforms as in \citep{Mercier:PoF:08}, the first limitation comes from the necessity to handle manually the selection of the spatial frequency range used to identify one of the modes, complicating an automatic procedure. The remaining mode is then deduced by subtracting the extracted mode to the original image. A second drawback is thus a sensitivity to noise, preventing the decomposition of modes having close spatial frequencies. Finally, the Hilbert transform  operation is based on spatial Fourier transformation that strongly reduces the spatial resolution at the boundaries which are specifically the regions we are interested in studying.

In this study, we propose to adapt the 2-D Variational Mode Decomposition (2-D VMD) \citep{Zosso2017} method to perform mode decomposition. This problem can be formulated as an inverse problem which consists  in extracting $J$ oscillating components (modes), denoted $(m_j)_{1 \leq j \leq J}$ with $m_j \in \mathbb{R}^{N_{x_s} \times N_{z_s}}$, where $N_{x_s} \times N_{z_s}$ is the number of points of the grid, from the observed data $d$ such that 
\begin{equation}
d= \sum_{j=1}^J m_j + \varepsilon,
\label{eq:sumModes}
\end{equation}
where $\varepsilon\sim \mathcal{N}(0,\sigma_n^2 I)$ models an additive Gaussian noise such as measurement noise.

In order to fit VMD formalism, the expression of $m_j$ provided in equation~\eqref{eq:sumModes} for the study of internal wave reflection corresponds to the specific case $J = 2$, even if the 2D-VMD approach is developed for a general $J\in \mathbb{N}_*$ value. For all the modes $m_j$, the spatial wave components, $k_{x_s,j}(x_s,z_s)$ and $k_{z_s,j}(x_s,z_s)$, are centered around the unknown spatial frequencies $(v_{x_s,j},v_{z_s,j})$, that are independent of the spatial coordinates.
The values of $(v_{x_s,j},v_{z_s,j})$ are thus close to the spatial average of the spectral content $k_{x_s,j} \in \mathbb{R}^{N_{x_s} \times N_{z_s}}$  (resp. $k_{z_s,j}$). At every location, $(x_s,z_s) \in \{1,...,N_{x_s}\} \times \{1,...,N_{z_s}\}$, the velocity field of the mode $j$ for the temporally filtered field can be estimated:
\begin{equation}
m_{j}(x_s,z_s)= a_j(x_s,z_s)\cos\left(v_{x_s,j}x_s +v_{z_s,j} z_s + \phi\right),
\end{equation}
where $a_j \in  \mathbb{R}^{N_{x_s} \times N_{z_s}}$ models the amplitude changes in space and $\phi$ is a phase term. 

The 2-D VMD aims at estimating jointly $(m_j)_{1 \leq j \leq J}$,  $(v_{x_s,j})_{1 \leq j \leq J}$ and $(v_{z_s,j})_{1 \leq j \leq J}$ by solving
{\small{\begin{multline}
 \underset{(m_j, v_{x_s,j}, v_{z_s,j})_{1 \leq j \leq J}}{\textrm{minimize}} \left \{ \norm{  d-\sum_{j=1}^J m_j  }^{2}\right.  \\
\left.+ \alpha \sum_{j=1}^J \Big\Vert D_{(x_s,z_s)} \big(m^{AS}_{j}(x_s,z_s)e^{-i(v_{x_s}x_s+v_{z_s}z_s)} \big) \Big\Vert^{2} \right \},
\label{eq:Reflection:Observation:VMD}
\end{multline}}}
where $D_{(x_s,z_s)}$ models the spatial discrete gradient operator and the coefficient $\alpha > 0$ denotes a regularization parameter allowing to adjust the bandwidth size of the filter. The 2D analytic signal $m^{AS}_{j}$ is the inverse Fourier transform of $\widehat{m}^{AS}_{j}$, which is defined  in the Fourier domain as

{\small{\begin{align}
\widehat{m}^{AS}_{j}(\nu_{x_s},\nu_{z_s}) = \left(1+\text{sign}(v_{x_s,j}\nu_{x_s}+v_{z_s,j}\nu_{z_s})\right) \widehat{m}_{j}(\nu_{x_s},\nu_{z_s}),
\end{align}}}
where $\widehat{{m}_j}$ is the Fourier transform of $m_j$. Note that the 2D analytic signal is chosen to set to zero one half-plane of the spatial frequency domain relatively to $(v_{x_s,j})_{1 \leq j \leq J}$ and $(v_{z_s,j})_{1 \leq j \leq J}$.

\subsection{Specificities to internal wave reflections}

We improved this model considering the specific properties of internal wave reflections:
\begin{itemize}
\item first, incident and reflected waves have different spectral behaviors. In particular, due to the focalisation effect close to criticality, the spectrum of the reflected wave is very compact horizontally but not vertically. Parameters $\alpha_{j,x_s}$ and $\alpha_{j,z_s}$ depending on the mode $j$ and the axis direction have been introduced, in order to separately adjust the along-slope and normal spectral compacity of each mode. 
\item Second, for the critical and near critical reflections, the reflected wave will stay in the proximity of the boundary. For this case we expect that the mode associated to the reflected wave will vanish far away from the slope. This information can be introduced through a penalty term $f_j(m_j )$, which acts as an indicator function $i_C(m_j )$ whose value is $0$ if $m_j \in C = \{ \mathbf{v} \in \mathbb{R}^{N_{x_s} \times N_{z_s}} | (\forall (x_s,z_s) \in \mathbb{S})\quad v(x_s,z_s)=0	\} $ and $+ \infty $ otherwise. For such a choice of the penalty $f_j$, we impose the component $m_j$ to be zero in the set of indexes $\mathbb{S}$, which is a chosen subset of $\mathbb{R}^{N_{x_s} \times N_{z_s}}$.
\end{itemize}
According to these remarks, we aim to solve:

{\small{\begin{align}
&  \underset{(m_j, {v}_{x_s,j}, {v}_{z_s,j})_{1 \leq j \leq J}}{\textrm{{minimize}}}     \Bigg \{    \sum_{j=1}^J f_j(m_j) + \theta \norm{ \sum_{j=1}^J m_j  - d }^{2}   \nonumber\\
 +& \sum_{j=1}^J  \alpha_{x_s,j} \Big\Vert D_{x_s} \big(u^{AS}_{j}(x_s,z_s)e^{-iv_{x_s,j}x_s}\big)_{(x_s,z_s)} \Big\Vert^{2}  \nonumber\\
 +& \sum_{j=1}^J  \alpha_{z_s,j}\Big\Vert D_{z_s} \big( u^{AS}_{j}(x_s,z_s)e^{-iv_{z_s,j}z_s}\big)_{(x_s,z_s)} \Big\Vert^{2}  \Bigg \},
\label{eq:Reflection:Observation:VMDnoniso}
\end{align}}}
where $D_{x_s}$ and $D_{z_s}$ denote respectively the discrete gradient operator along the $x_s$ and $z_s$ component. The parameters $\alpha_{x_s,j}$ and $\alpha_{z_s,j}$ which allow for the adjustement of the selectivity for each mode and component are chosen positive. The parameter $\theta$ permits to adjust the attachment of the decomposition to the data $d$. In \citep{Schmitt:EUSIPCO:15}, we proposed an efficient algorithmic scheme based on alternating proximal algorithm to provide a local minimizer of this minimization problem. One could note that such alternating minimization strategies are applied for various fields in inverse problems:~\citet{Attouch2010,Bolte2010,Foare2019}. For a further and detailed discussion of the impact of the parameters and their choice see \citet{Schmitt:EUSIPCO:15}. In addition, in this same reference the authors perform a comparison between the Hilbert classical decomposition method, the basic 2D-VMD model (Eq.~\ref{eq:Reflection:Observation:VMD}) and the proposed 2D-VMD-prox decomposition model (Eq.~\ref{eq:Reflection:Observation:VMDnoniso}), the latter, developed for internal wave near critical reflections, showing a better performance to accurately isolate both wave modes and resolve the wave at the boundaries of the field of view.

\section{\label{sec:appli}Application of VMD to internal wave reflection}

In this work the observed data $d$ is the along-slope component of the velocity. In a first step in order to adjust the regularization parameters of the method we will use synthetic data before applying these adjusted parameters to decompose into modes the experimentally measured quantity. We will use respectively the terms $mode$ 1 and $mode$ 2 for the field associated to the incident and the reflected wave.

\subsection{\label{subsec:synthetic}Synthetic field of an internal wave critical reflection}

\begin{figure}[h]
\centering
\includegraphics[width=7.55cm]{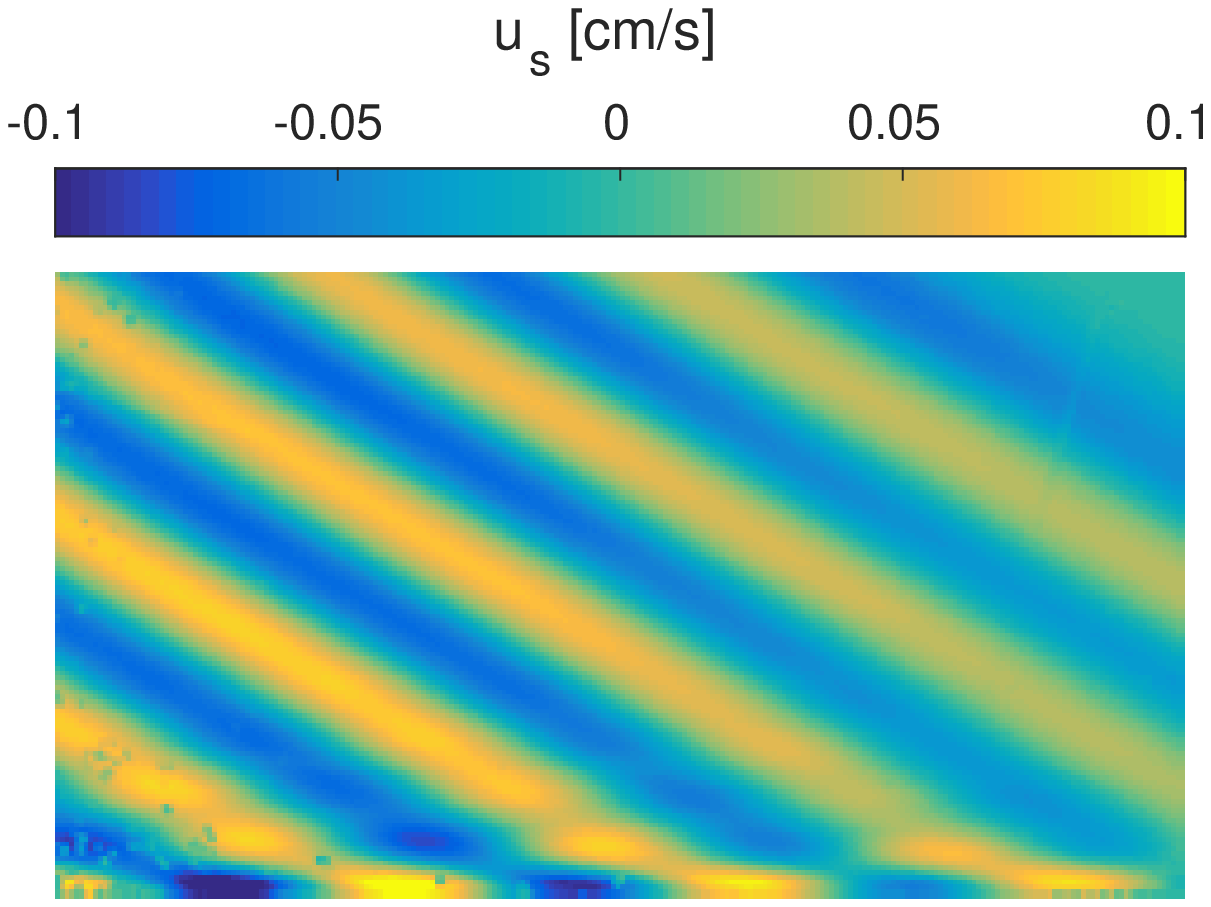} 
\includegraphics[width=7cm]{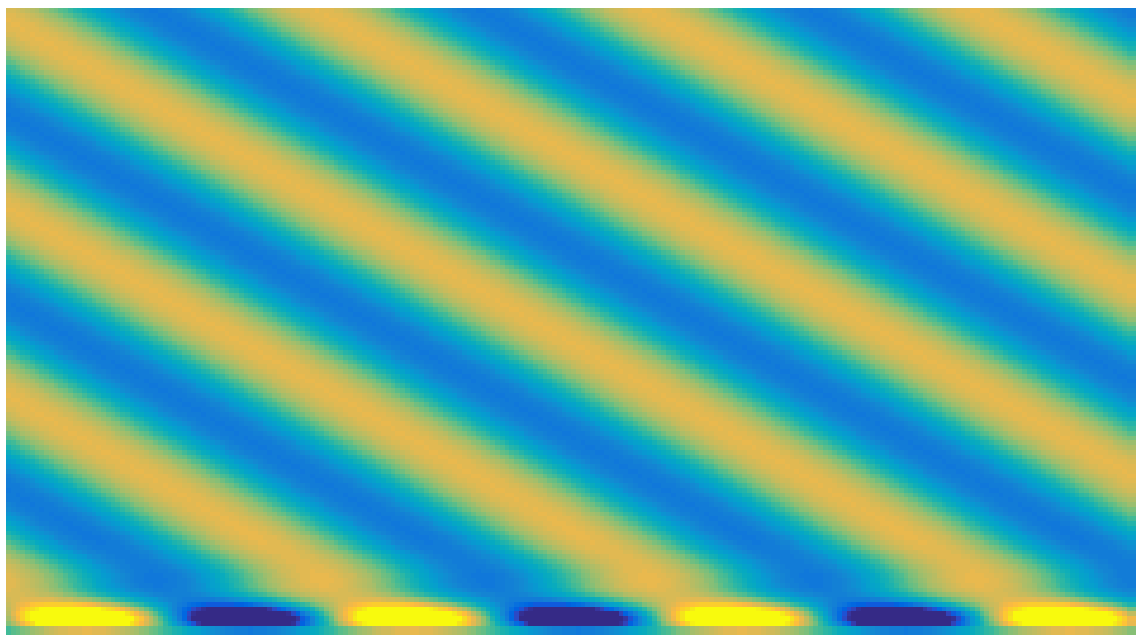} 
\includegraphics[width=7cm]{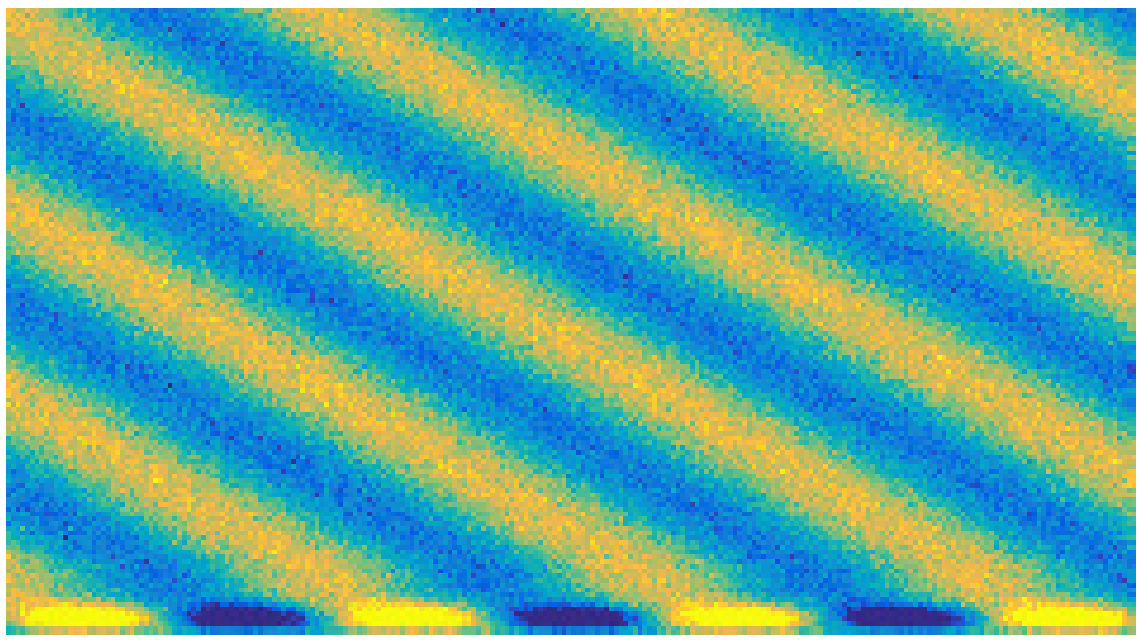}
\caption{Field of view of the along-slope component of the velocity for a critical reflection corresponding to ``exp3'' (top). A synthetic image $U_{\mathrm{syn}}$, based on experimental parameters of ``exp3'' without ($C_n = 0$) (center) and with added noise ($C_n = 0.3$) (bottom). The axes are the same as in Figure~\ref{figure3} (Left) yet they are not represented for the sake of clarity since the main message here is the general aspect of the velocity field.}
\label{figure4}
\end{figure}

The validation of the decomposition method is performed using a synthetic image of an internal wave critical reflection. The synthetic image of the along-slope component of the velocity is built by adding an incident synthetic wave, a reflected synthetic wave, and superimposed additive noise, formally $d = U_{\mathrm{syn}} = U^{\mathrm{inc}}_{\mathrm{syn}} + U^{\mathrm{refl}}_{\mathrm{syn}} +\varepsilon$. For each experiment we produced a synthetic image that is constructed using the physical parameters of the experiment. The incident wave is a plane wave with a known velocity and angle of propagation. For the reflected wave we use the solution derived  by~\cite{Dauxois:JFM:99}, refered to as the D-Y model in the following. The viscous expression for the along-slope component of the reflected wave at criticality is given by:

{\small{
\begin{align}
U^{\mathrm{refl}}_{\mathrm{syn}}&(x_s,z_s,t)  = \nonumber\\ &
\frac{ \mathrm{Re}^{1/3} U}{2\sqrt{3}\sqrt{(N/\omega_0)^2-1}}e^{-z_s/(2a)} \cos(k_{x_s} x_s - \omega_0t) \nonumber\\ 
&\left[\sin \left( 
\frac{\sqrt{3}z_s}{2a}+\pi/3 
\right)\right.  
-\left. \sqrt{3}\cos\left(\frac{\sqrt{3}z_s}{2a}+\pi/3\right)\right] ,
\label{theoryDauxois}
\end{align}
}
}
where Re is the Reynolds number associated to the incident wave, $U$ is the maximum velocity of the incident wave and $a$ a characteristic length given by
\begin{equation}
a = \left(\frac{\nu+\kappa}{(4\omega_0 k\left[1-(\omega_0/N)^2\right]}\right)^{1/3},
\end{equation}
where $\nu$ is the kinematic viscosity, $\kappa$ the salt diffusivity and $k = \norm{\mathbf{k}}$. One can observe in equation~(\ref{theoryDauxois}) that the intensity of the reflected wave is related to the intensity of the incident wave. The intensity of the incident wave is obtained from the experimental observations and is computed in the region far from the slope to avoid the component related with the reflected wave which is confined to the surrounding of the slope. The angle of propagation of the incident wave is inferred through the relation $\omega_0/N = \sin \beta$. The added noise has the form $\varepsilon\sim \mathcal{N}(0,\sigma_n^2 I)$ where $\sigma_n = C_n \text{std}(U_{\mathrm{syn}})$ and $C_n$ is a multiplicative coefficient that modulates the intensity of the noise. In Figure~\ref{figure4} we show a snapshot of an experimental measurement of the along-slope component of the velocity (top), a synthetic image based on the same experiment (center) and at the bottom the synthetic image with added noise ($C_n = 0.3$). The main features of the experimental observations are captured in the synthetic image: the incident wave angle of propagation, wavelength and intensity are in agreement; the reflected wave is located in the surroundings of the slope and spatial patterns of the variations of intensity are well represented. The lower order decay of the intensity of the wave as it distances from its source is not contained in the synthetic field. 

\begin{figure}[h]
\centering
\includegraphics[width=\columnwidth]{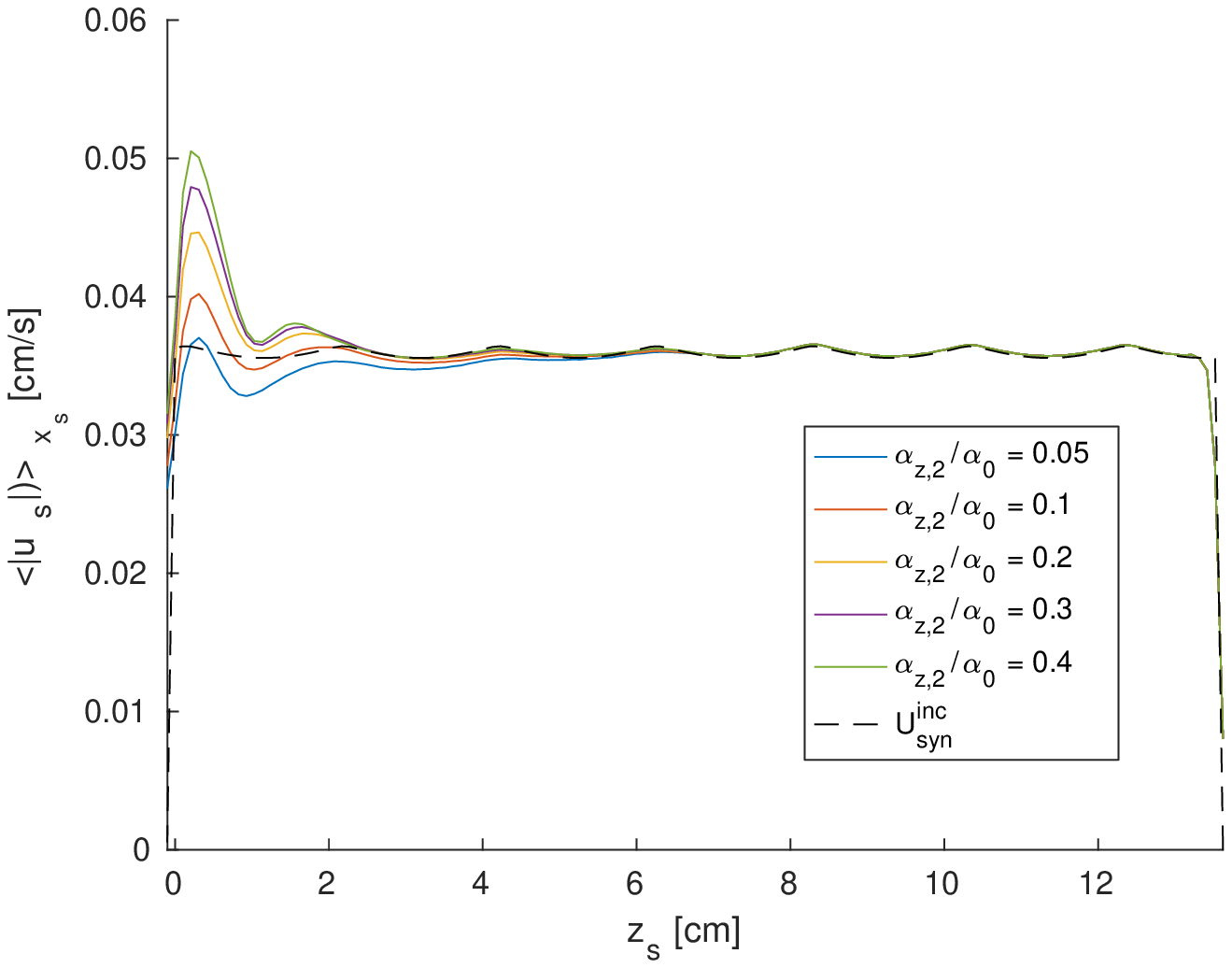} 
\includegraphics[width=\columnwidth]{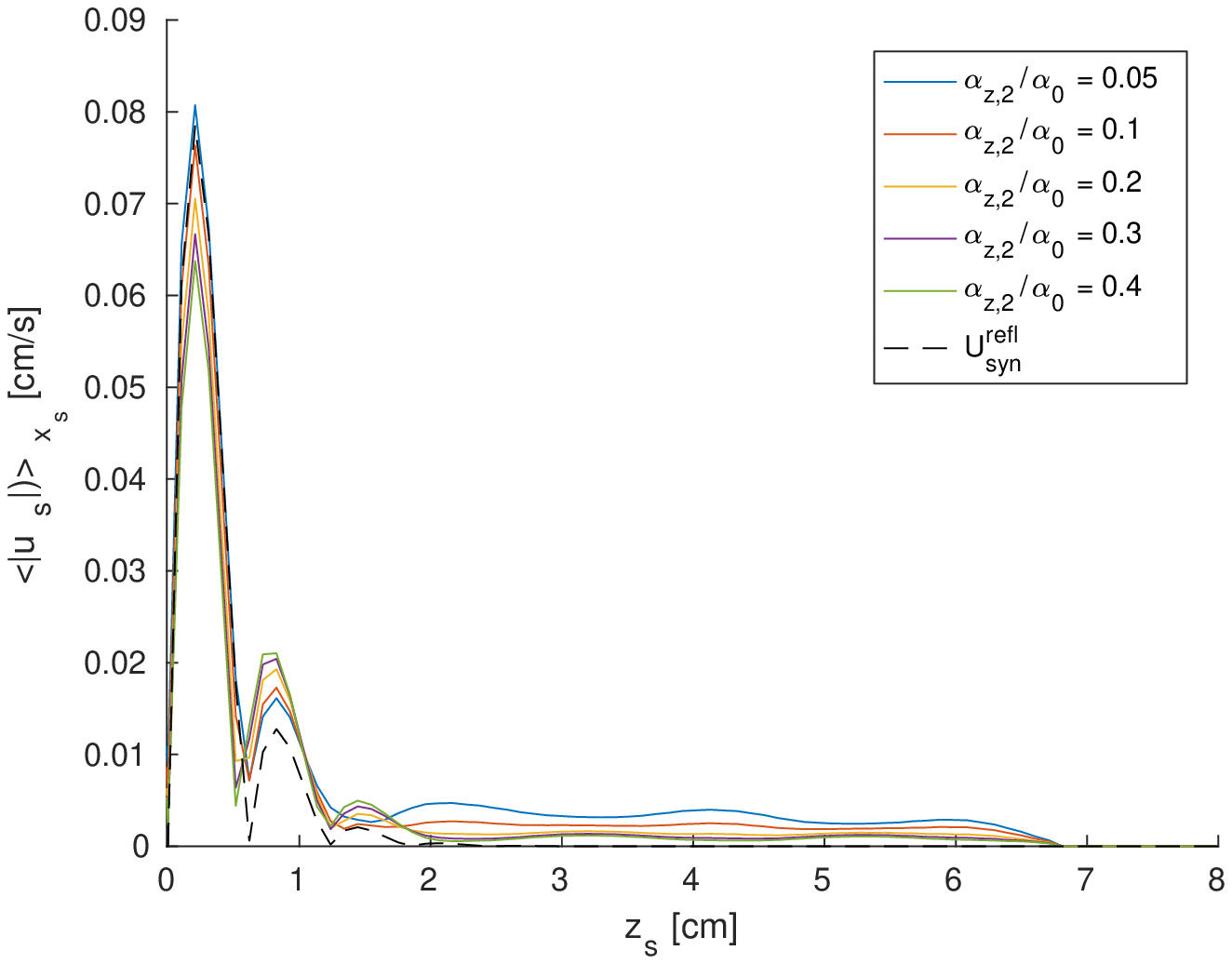} 
\caption{Profile along $z_{s}$ of the absolute value of $u_s$ averaged along the coordinate $x_s$ for \textit{mode 1} (top) and \textit{mode 2} (bottom) obtained from a synthetic image $U_{\mathrm{syn}}$. $U^{\mathrm{inc}}_{\mathrm{syn}}$ and $U^{\mathrm{refl}}_{\mathrm{syn}}$ are represented in dashed black lines. The profiles are shown for different values of $\alpha_{z,2}/\alpha_0$. The profiles of $mode$ 2 are shown in a slightly zoomed region near the slope ($z_s = [0,8]$ cm) in order to highlight the region where the reflected wave is located.}
\label{figure5}
\end{figure}

\subsection{VMD parameters selection}

The 2D-VMD-prox decomposition method is applied to the synthetic image of the along-slope component of the velocity field $U_{\mathrm{syn}}$ described in section~\ref{subsec:synthetic} for a given temporal phase choice. The two output $modes$ are compared with the components of the synthetic image $U^{\mathrm{inc}}_{\mathrm{syn}}$ and $U^{\mathrm{refl}}_{\mathrm{syn}}$.
In order to better assess the result of the 2D-VMD-prox decomposition method we use vertical profiles of the 2D images. To do that, we compute the along-slope spatial average (direction $x_s$) of the absolute value of the along-slope component of the velocity field ~$<|u_s|>_{x_s}$. In Figure~\ref{figure5}(Top) (resp. Bottom) $<|u_s|>_{x_s}$ as a function of $z_s$ is shown for the $mode$ associated to the incident and reflected wave respectively. Both, the incident and reflected $modes$ are well differentiated showing no overlap of one $mode$ over the other. In addition, they are in good agreement with the synthetic image. For the $mode$ corresponding to the incident wave the profile reproduces perfectly the synthetic profile some distance away from the slope. The small oscillations of the amplitude of the profile are a consequence of performing an average in the direction of $x_s$ over a distance that is not an entire multiple of the wavelength of the wave. For the $mode$ corresponding to the reflected wave the profile shows that the strong increase and decrease of the amplitude in the surroundings of the slope ($z_s \simeq 0$) are captured, as well as the width of these peaks. At the extreme right of the profile, the penalisation acts over the reflected $mode$ and the amplitude of the profile is imposed to be equal to zero. Between the peaks near the slope and the penalisation region, the amplitude of the profile is steady and even though it struggles to go strictly to zero, its amplitude is fairly small.

There are, however, some discrepancies between the decomposed $modes$ and the synthetic images. They are found in the region near the slope, where the overlap of the two $modes$ is major. To minimize these differences we optimize the relation between the parameters $\alpha_{x,j}$ and $\alpha_{z,j}$. These parameters permit to take into account the spectral compactness of each mode in each direction.

\begin{figure}[h]
\centering
\includegraphics[width=\columnwidth]{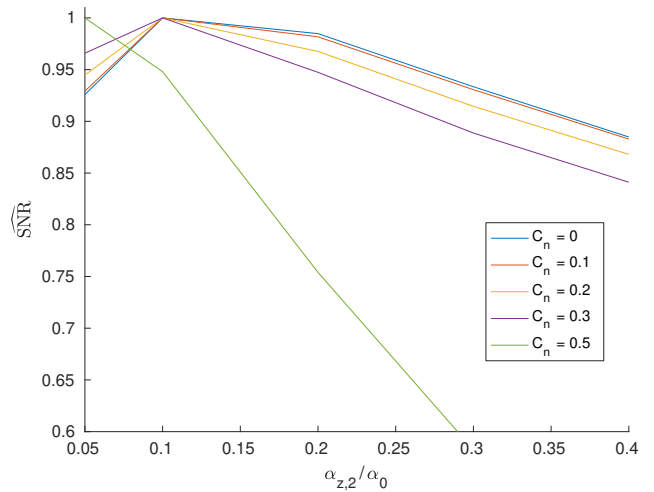}
\caption{$\widehat{\mathrm{SNR}}$ as a function of $\alpha_{z,2}/\alpha_0$ for several values of noise standard deviation (modulated by $C_\mathrm{n}$) quantifying the comparison between the extracted $mode$ associated to the reflected wave and the synthetic reflected wave constructed following experimental parameters of ``exp3'' and D-Y model~(Eq.~\ref{theoryDauxois}).}
\label{figure6}
\end{figure}

Near the critical angle, a small departure from criticality corresponds to a large variation of $k^{\mathrm{refl}}_{z_s}$. The conservation of $k_{x_s}$ and $\omega_0$ in the reflection leads indeed, in the linear theory of internal wave reflections, to the following relation between the incident and reflected vertical wavenumbers: $k^{\mathrm{refl}}_{z_s} = k^{\mathrm{inc}}_{z_s} \tan (\beta +\gamma)/\tan (\beta -\gamma)$. Therefore, a small deviation of $|\beta-\gamma|$ from $0$ implies a large variation of $k^{\mathrm{refl}}_{z_s}$. On the other hand, for the angles $\beta$ and $\gamma$ explored in this work $\mathcal{O}(k^{\mathrm{inc}}_{x_s})=\mathcal{O}(k^{\mathrm{inc}}_{z_s}$). Therefore, we set the parameters $\alpha_{x,1} = \alpha_{x,2} = \alpha_{z,1}$ to a given value of $\alpha_{0} = 100$, and we use the parameter $\alpha_{z,2}$ associated to the wavenumber $k^{\mathrm{refl}}_{z_s}$ as a selectivity parameter to optimize the out-coming $modes$. 
In Figure~\ref{figure5} the profiles of the incident and reflected $modes$ are plotted for several values of $\alpha_{z,2}/\alpha_0$.
The variation of the selectivity parameter $\alpha_{z,2}$ modifies primarily the profile of the decomposed $modes$ in the region near the slope where deviations between the synthetic image and the decomposed $modes$ can be the largest. 
To select the optimal ratio between $\alpha_{z,2}$ and $\alpha_0$ we use the signal-to-noise ratio defined as

\begin{align}
\text{SNR}(u^{\mathrm{true}},u^{\mathrm{estimated}}) = 20 \log_{10}\left(\frac{u^{\mathrm{true}}}{u^{\mathrm{estimated}}-u^{\mathrm{true}}}\right).
\end{align}
The SNR provides a quantitative evaluation of the similitude of each $mode$ with the synthetic data when varying $\alpha_{z,2}$ and $C_n$. It is shown in Figure~\ref{figure6} the plot of $\widehat{\text{SNR}}=\text{SNR}/\text{max(SNR)}$ as a function of $\alpha_{z,2}/\alpha_0$ for each value of $C_n$ explored. $\widehat{\text{SNR}}$ is obtained by comparing between $U_{\mathrm{syn}}^{\mathrm{refl}}$ and the $mode$ associated to the reflected wave recovered from the 2D-VMD-prox decomposition. This procedure is performed for different degrees of noise standard deviation $C_n$ in order to measure the robustness of the results. The highest value of $\widehat{\text{SNR}}$ corresponds to the value of $\alpha_{z,2}/\alpha_0 = 0.1$, for a noise coefficient between $0$ and $0.3$. For a $C_n$ equal to $0.5$ or higher (not shown) we observe that the method struggles to correctly identify the two modes.

The selectivity parameter  $\alpha_{z,2}$ that maximizes the $\widehat{\text{SNR}}$ when applied to the synthetic images is then used for the decomposition of the associated experimental images. 

\subsection{\label{sec:Reflection:Results}Decomposition of experimental measurements}

\begin{figure*}[h]
\centering
\hspace{1.5cm} \includegraphics[width=5.9cm]{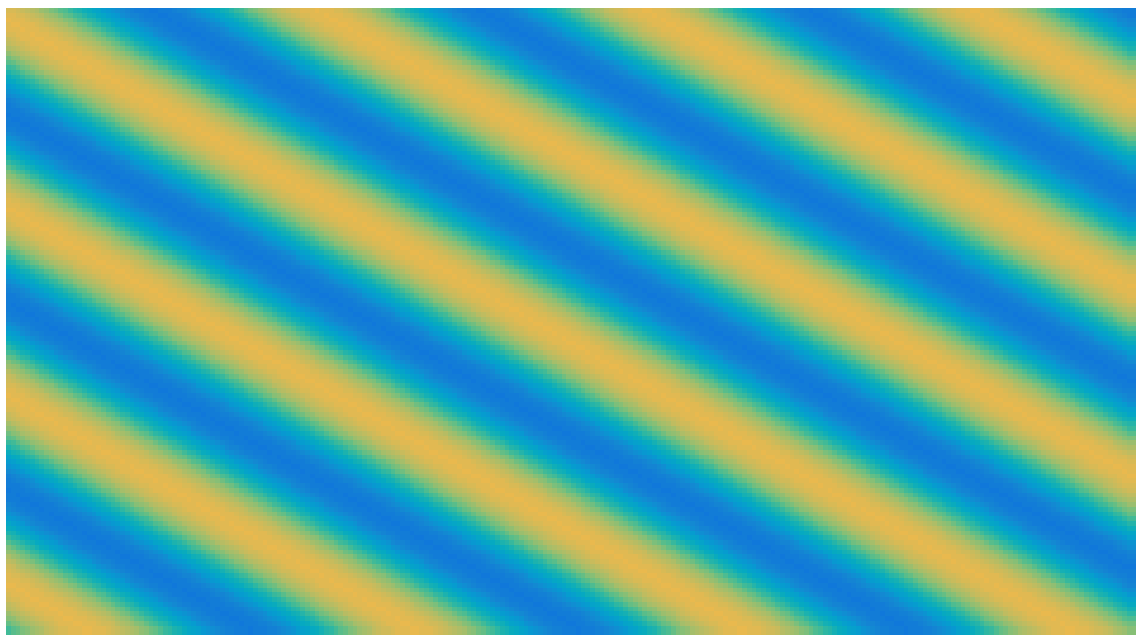}
\includegraphics[width=5.9cm]{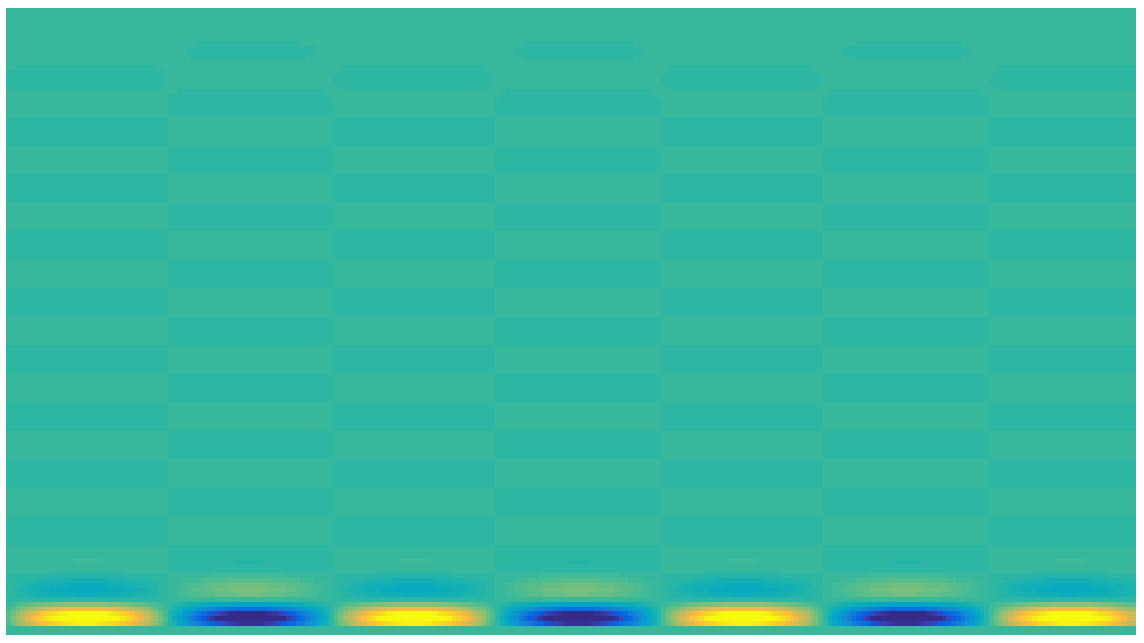}
\raisebox{-0.1cm}{\includegraphics[width=3.6cm]{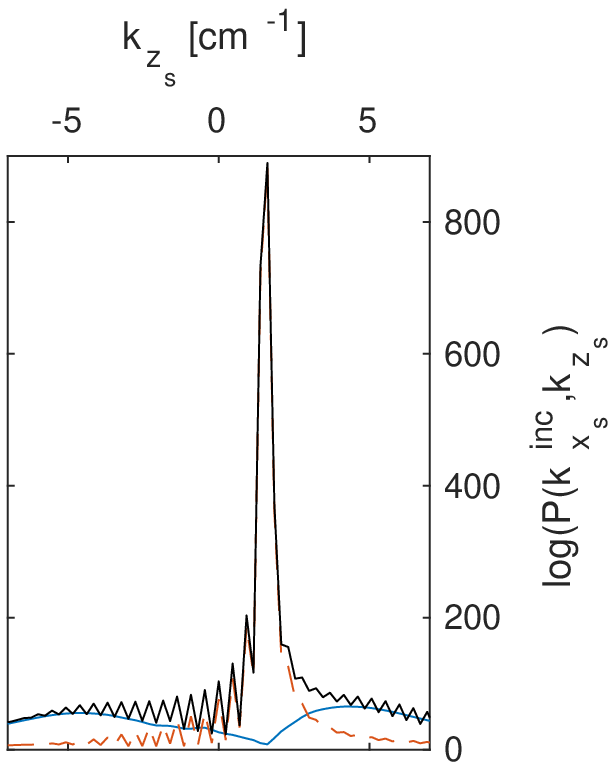}}\\
\raisebox{-0.12cm}{\includegraphics[width=7.5cm]{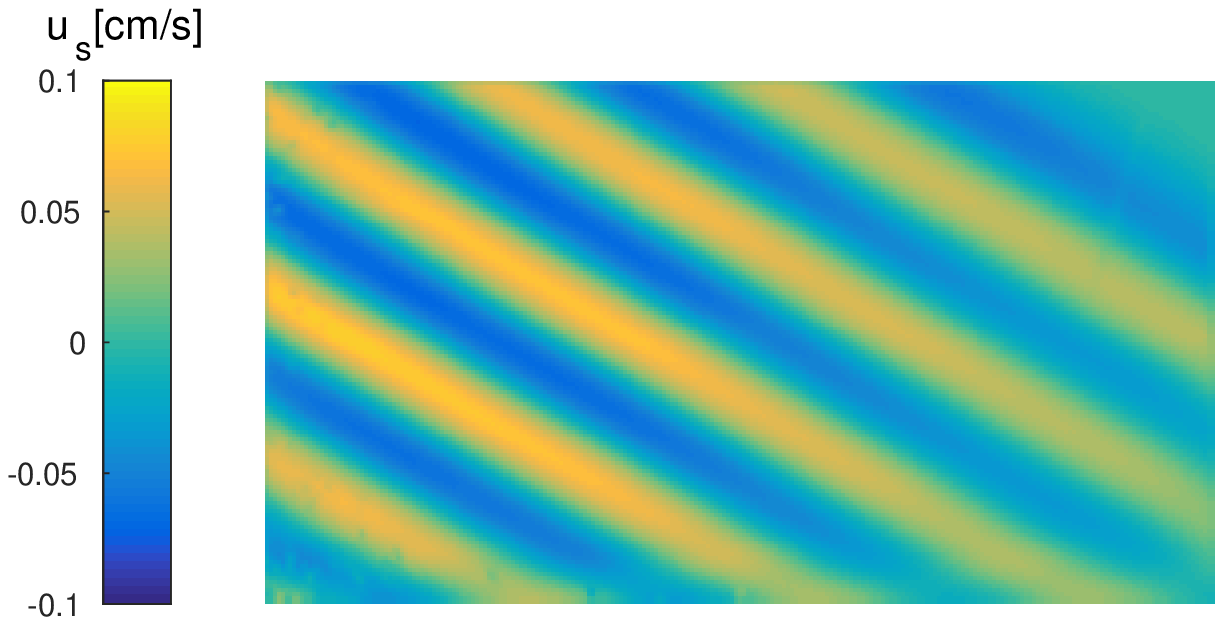}}
\includegraphics[width=5.9cm]{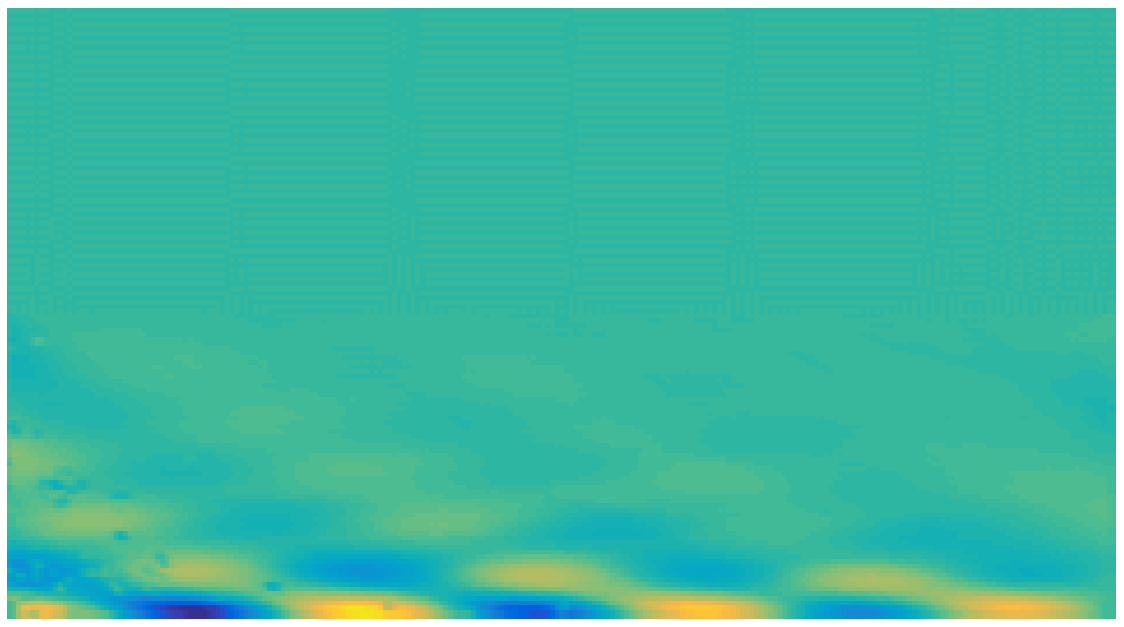}
\raisebox{-0.1cm}{\includegraphics[width=3.6cm]{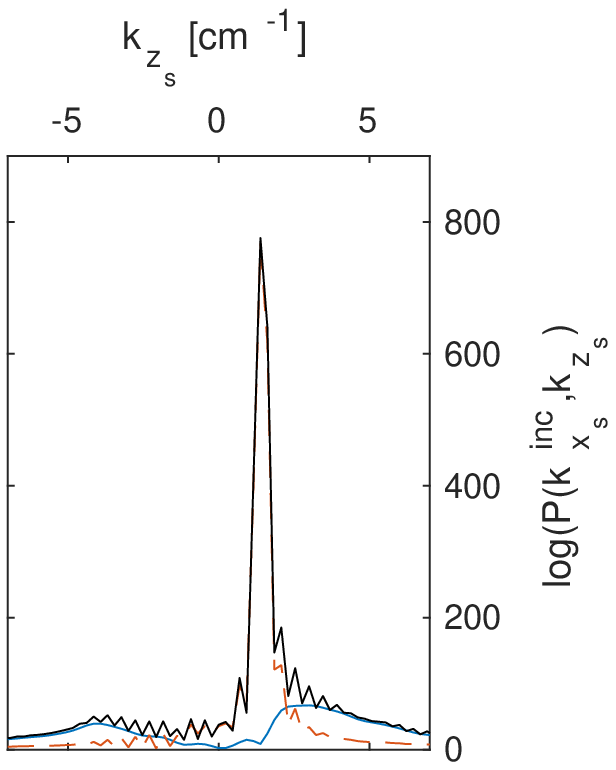}}
\\
\hspace{2cm}\textit{Mode 1} \hspace{5cm} \textit{Mode 2} \hspace{3.5cm}\textit{Spectra}

\caption{2D-VMD-prox decomposition method applied to the along-slope component of the velocity of a critical reflection for a synthetic field $U_{syn}$ (first row) and for the corresponding experimental field \textit{input} (second row). \textit{mode~1} (first column) and \textit{mode~2} (second column) are associated respectively to the incident and the reflected wave. The third column shows the spatial spectrum profile along $k_{z_s}$ at $k_{x_s} = k^{\mathrm{inc}}_{x_s}$ for the data field $d$ (black), \textit{mode 1} (red) and \textit{mode 2} (blue). The experiment corresponds to case ``exp3''. For the 2D images the axes are the same as in Figure~\ref{figure3} (Left) yet they are not represented for the sake of clarity. }
\label{figure7}
\end{figure*}

Figure~\ref{figure7} displays the image of the 2D-VMD-prox decomposition $modes$ obtained when applied over the synthetic field associated to ``exp3'', $d = U_{syn}$ (first row) and the corresponding experimental measurements \textit{d = input}, (bottom row). The first column shows the \textit{mode 1} associated to the incident wave, the second column shows the  \textit{mode 2} associated to the reflected wave and the third column shows the spectral profile at $k_{x_s} = k^{\mathrm{inc}}_{x_s}$ of the vertical spectral content showing the relative contributions of each $mode$. The overlap between the reflected and the incident wave in the spatial spectrum is observed in both the experimental and in the synthetic results and is predominant for the values around $k_{z_s} \sim 2$ cm$^{-1}$. The spectral overlap of these two waves is an indication of the intrinsic difficulty that the critical reflection presents to isolate incident and reflected wave. For the experimental results the mode decomposition is well achieved and there is practically no overlap between the incident and the reflected waves on each $mode$ obtained by the 2D-VMD-prox decomposition. The $mode$ 1 associated to the incident wave and the synthetic incident wave present very similar features which is not a surprise as the synthetic image is constructed using measurements of the experimental incident wave. In the case of the reflected wave, the oscillations of the pattern in the region near the slope are well captured for the synthetic reflected wave, both in wavelength and in width ($z_s$ direction). The intensity of the oscillations of the experimental reflected wave are weaker than for the synthetic reflected wave, which could be a consequence of the experimental limitations to achieve the exact critical reflection ($\gamma = \beta$). 

\begin{figure}[h]
\centering
\includegraphics[width=8cm]{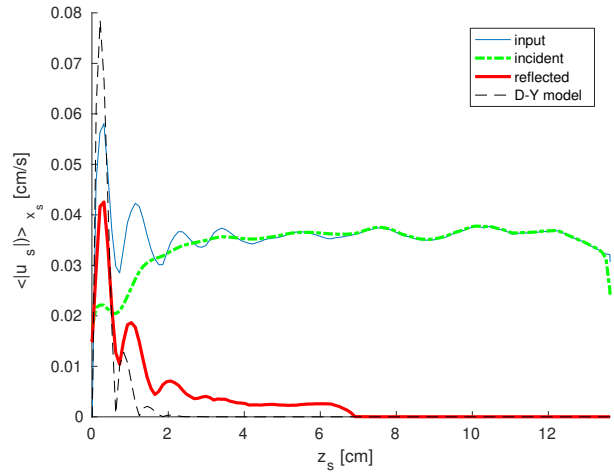}
\caption{Profile of the incident and the reflected wave of ``exp3'' obtained using the 2D-VMD-prox decomposition method with $\alpha_{z,2}/\alpha_0 = 0.1$. The Dauxois-Young model profile (Eq.~\ref{theoryDauxois}) for the reflected wave is plotted in black dashed line.}
\label{figure8}
\end{figure}

For a quantitative comparison of the reflected wave extracted from the decomposition with the D-Y theory, we use the profile $<|u_s|>_{x_s}$. In Figure~\ref{figure8} the profile $<|u_s|>_{x_s}$ as a function of $z_s$ is shown for the $input$ data field and for the two $modes$ obtained by the 2D-VMD-prox method of ``exp3''. The $input$ profile is fully represented by the $mode$ 1 associated to the incident wave far from the slope. Near the slope, the profile of the incident wave presents a decrease of its amplitude. This could be partially related to the fact that the slope is on average further away from the wave source than the rest of the field of view and therefore, the viscous decay of the incident wave is larger near the slope. Note that we performed the decomposition over the temporally filtered experimental velocity field at different instants, i.e., different temporal phase term and we observed that the results shown in Figure~\ref{figure8} are not affected by the selection of the phase.
In black dashed line, the profile of the D-Y model is shown for the reflected wave, corresponding to the expression of equation~\ref{theoryDauxois}. The D-Y model describes successfully the width of the amplitude-oscillations of the reflected wave near the slope, however, the relative amplitude of these oscillations is not completely captured by the model. In particular for the first amplitude peak (starting from the slope, $z_s =0$) the model overestimates by almost a factor 2 the amplitude of the first peak of the reflected wave. This overestimation may be related to the fact that this model represents the singular case of $\gamma \equiv \beta$, which is extremely complicated to achieve (or even get very close) when performing experiments. We observed that the overestimation of the first peak by the model is present in all the experiments performed in this work, which would tend to support this explanation. 

\begin{figure}[h]
\centering
{\includegraphics[width=8cm]{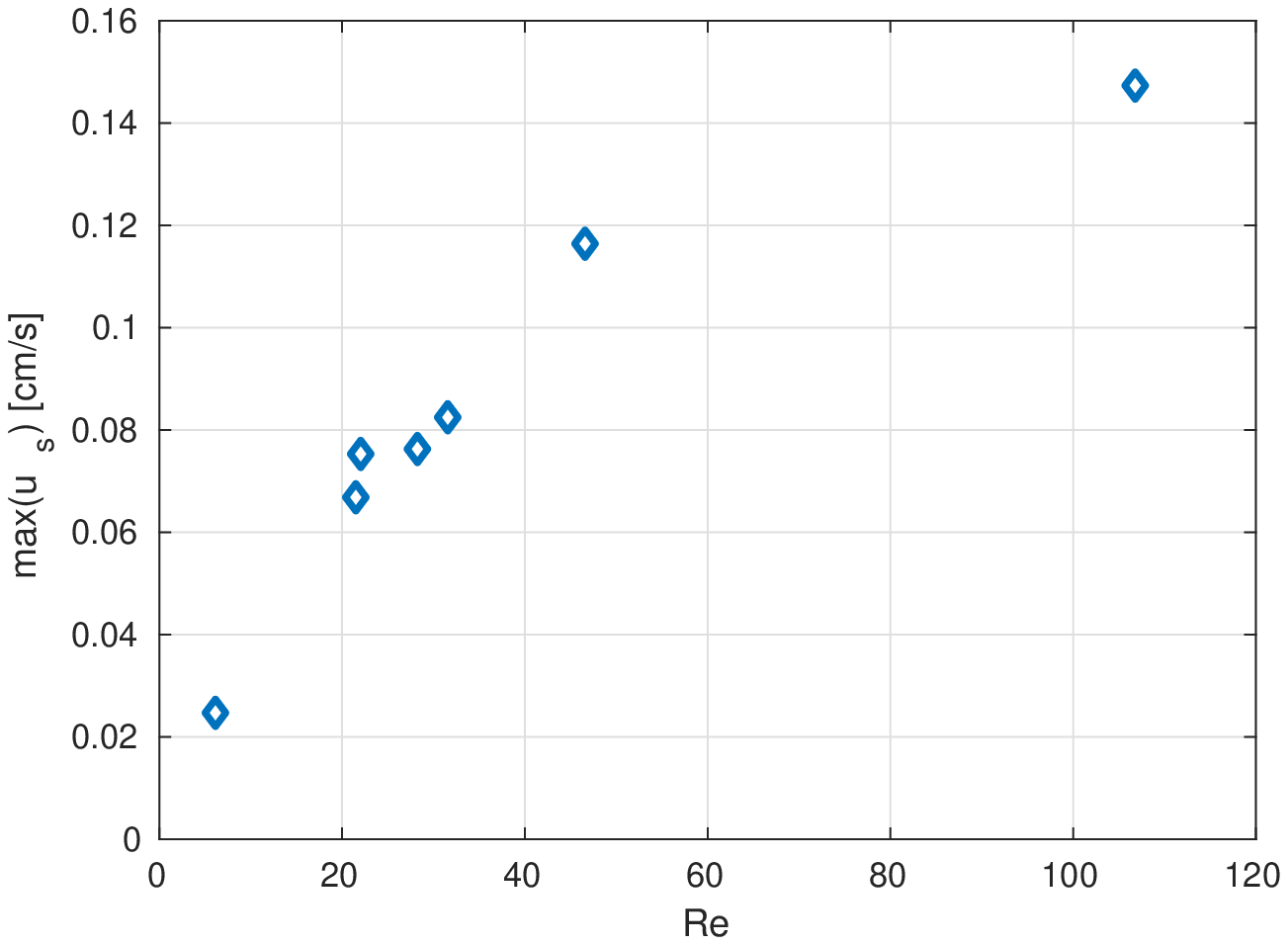}} 
{\includegraphics[width=8cm]{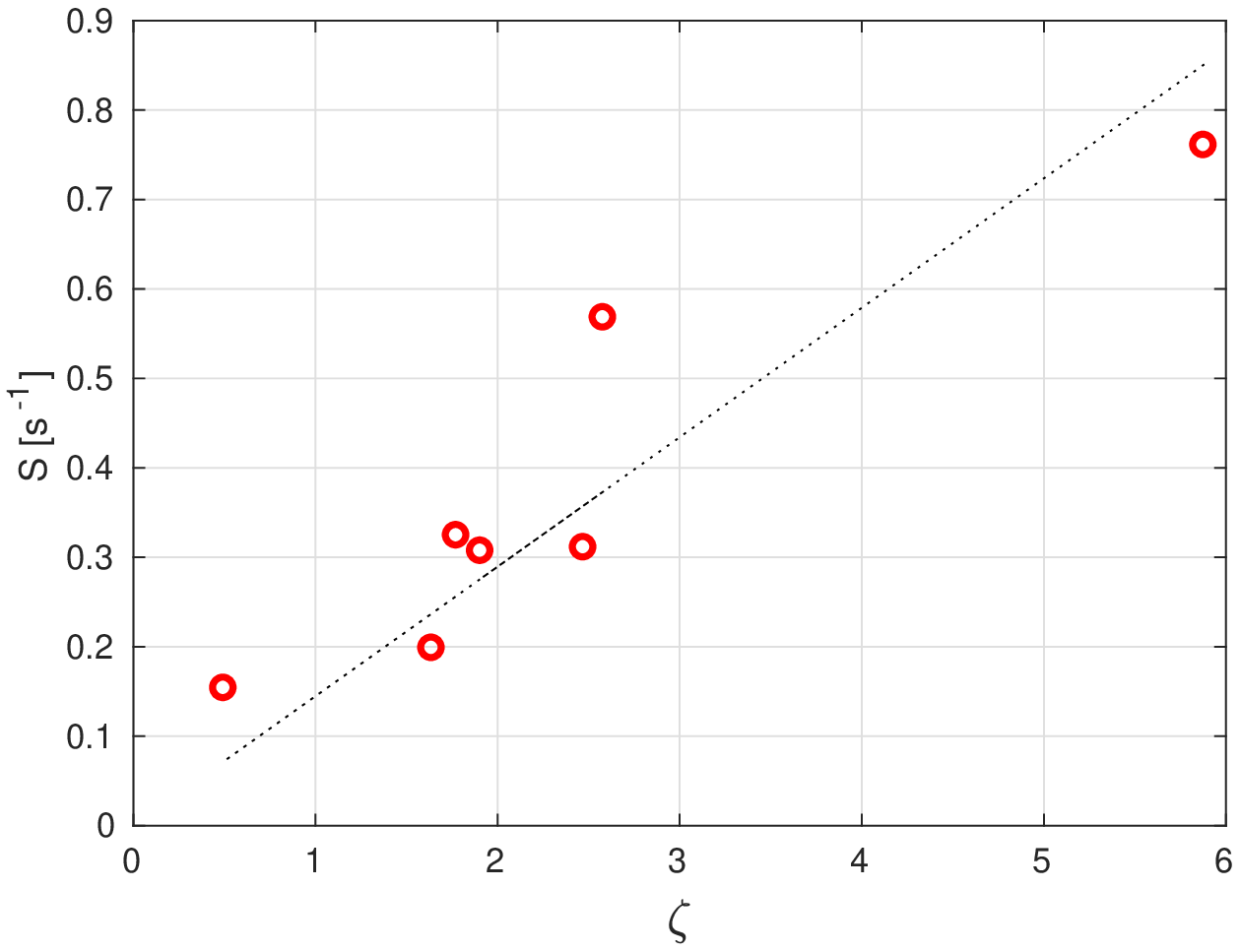}} 
\caption{(Top) $\max(u_s)$ for the \textit{mode 2} associated to the reflected wave obtained for the most critical reflection of each experiment as a function of the Reynolds number. (Bottom) Shear stress, $S = \max(u_s)/\delta$ as a function of the parameter $\zeta$. A linear regression is plotted in black dashed line.}
\label{figure9}
\end{figure}

In Figure~\ref{figure9} (top), the maximum value of the reflected wave is shown as a function of the Reynolds number for all experiments (see table \ref{table:reflection:system:exp}). The maximum value of the velocity of the reflected wave, as expected, increases as the value of the velocity of the incident wave increases. Using the value of the maximum velocity at the first peak $\max(u_s)$ and the half-width of this peak (defined as the distance $\delta$ of the maximum from the slope, since the peak falls off to zero at the slope) we can give a first order estimate of the shear rate $S = \max(u_s)/\delta$ produced near the slope. The first order estimate of the shear rate $S$ using the D-Y model through equation~\ref{theoryDauxois} shows that the shear is proportional to the parameter $\zeta=U (\lambda^{\mathrm{g}}/N)^{1/3} \nu^{-2/3}$ where $U$ is the maximum velocity of the incident wave. In Figure~\ref{figure9} (bottom), $S$ is shown as a function of this parameter $\zeta$. A linear fit (dotted line) helps to identify the linear relation between $S$ and $\zeta$. This is an indication that regardless of discrepancies between the D-Y model and the experimental data in the prediction of the maximum value of the reflected wave (either by an overestimation of the model or due to the difficulty to achieve an experimental critical reflection), the behavior of the first order shear rate $S$ can be collapsed to a predictable behavior as a function of a combination of the experimental parameters $U$, $\lambda^{\mathrm{g}}$, $N$ and $\nu$ suggested by the D-Y theory.

\section{\label{Conclusions}Conclusions} 

In this study we present high resolution observations of internal waves critical reflection and a method that successfully achieves to isolate the incident and the reflected wave involved in the process. We present a mode decomposition method adapted specially for internal wave critical reflection which is tested over synthetic and experimental images. This method has been developed to handle the failure of other methods to correctly isolate the incident and reflected waves involved in the physical process. The technical challenges presented by a critical reflection have been taken into account in the developement of the decomposition method 2D-VMD-prox: both waves have the same temporal frequency $\omega_0$; both waves have the same spatial frequency component along the slope coordinate $k^{\mathrm{inc}}_{x_{\mathrm{s}}} = k^{\mathrm{refl}}_{x_{\mathrm{s}}}$; the reflected wave is confined in a narrow region of the field of view; the wavelengths of the waves are only one order of magnitude smaller than the size of the image; and finally, the region of most physical interest is located at the boundary of the images.

The synthetic images of a critical reflection allowed for testing and tuning of the decomposition method in order to apply the method to our experimental measurement and optimize the decomposition of the incident and reflected wave. 
The reflected wave is compared with the D-Y model for critical reflections \citep{Dauxois:JFM:99}, which takes into account non-linearities and viscosity in the process. The latter is of dominant importance for experimental conditions and is not taken into account by most models. The D-Y model correctly describes the pattern of velocity intensity of the reflected wave in the region near the slope in both wavelength and width, nevertheless, the predicted amplitude of the velocity for the first peak of the reflected wave is higher than observed. This quantitative discrepancy can be associated to the experimental limitation to achieve exact critical reflection, for which case the amplification of the velocity of the reflected wave is the largest. 

A range of experimental parameters is covered in the experiments presented in this work, and for all these experiments the D-Y model correctly describes the tendency of the first order shear rate $S$, produced by the reflected wave near the slope. 
The shear produced by internal wave critical reflections is tightly related to the resuspension and transport of sediment in oceanic conditions. The sediment transport is modeled using the Shield dimensionless number $\Theta = S \mu /((\rho_{\mathrm{p}} - \rho)g d_{\mathrm{p}})$ where $\rho$ and $\rho_{\mathrm{p}}$ are respectively the density of the fluid and of the sediment, $d_{\mathrm{p}}$ is the typical size of the sediment and $\mu$ is the dynamic viscosity. Although the range of Re number explored in this work is only of two decades, obtaining a tendency of the shear rate $S$ with respect to the measurable observation parameters $U$, $\lambda^{\mathrm{g}}$, $N$ and $\nu$ may allow to identify the conditions for which bedload transport of particles in the ocean is achieved. The erosion and transport of particles occurs when the threshold Shields number $\Theta_{\mathrm{th}} \approx 0.12$ is exceeded~\citep{Ouriemi2007}.

The critical reflection is produced when $\beta = \gamma$, nevertheless, if $\beta = \gamma^+$ the reflected wave propagates up-slope and if $\beta = \gamma^-$ the reflected wave propagates down-slope. For a well defined angle of propagation of the incident wave $\beta$, this upward or downward propagation of the reflected wave does not prevent the 2D-VMD-prox method to successfully isolate incident and reflected waves. Nevertheless, for some experimental conditions and almost always in oceanic conditions, the spectral content of the internal waves involves a range of frequencies and therefore of angles of propagation. In these cases, it is possible to produce two reflected waves simultaneously (upward and downward). We have tested that the 2D-VMD-prox decomposition method can be used for separating more than two $modes$. In the present study, however, we chose to use only the two $mode$ decomposition given the difficulties to set the experimental conditions close enough to the critical configuration (within less than $1^{\circ}$). In addition, we wanted to reduce as much as possible the number of parameters involved in the decomposition. We therefore leave the 3-mode decomposition to future studies. 

The 2D-VMD-prox method was tested in the most challenging conditions. 
For experiments involving a larger number of wavelengths within the field of view and a higher resolution of the velocity near the slope, an improvement of the performance of the decomposition method can be expected. 

The \textsc{Matlab} toolbox containing the 2D-VMD-prox algorithm and examples of experimental internal wave critical reflection measurements are made publicly available. 

\begin{acknowledgements}
We thank T. Dauxois for insightful discussions. This work has been partially supported by the ONLITUR grant ANR-2011-BS04-006-01 and achieved thanks to the resources of PSMN from ENS de Lyon.
\end{acknowledgements}

\bibliographystyle{ExpFluids}       

\bibliography{manuscript}   

\begin{thebibliography}{45}
\providecommand{\natexlab}[1]{#1}
\expandafter\ifx\csname urlstyle\endcsname\relax
  \providecommand{\doi}[1]{doi:\discretionary{}{}{}#1}\else
  \providecommand{\doi}{doi:\discretionary{}{}{}\begingroup
  \urlstyle{rm}\Url}\fi

\bibitem[{Attouch et~al.(2010)Attouch, Bolte, Redont, and
  Soubeyran}]{Attouch2010}
Attouch H, Bolte J, Redont P, Soubeyran A (2010) {An approach based on the
  Kurdyka- Lojasiewicz inequality}.
\newblock Mathematics of Operations Research 35:438--457

\bibitem[{Bogucki et~al.(1997)Bogucki, Dickey, and Redekopp}]{Bogucki:JPO:97}
Bogucki D, Dickey T, Redekopp LG (1997) {Sediment resuspension and mixing by
  resonantly generated internal solitary waves}.
\newblock {Journal of Physical Oceanography} 27({7}):{1181--1196}

\bibitem[{Bolte et~al.(2010)Bolte, Combettes, and Pesquet}]{Bolte2010}
Bolte J, Combettes PL, Pesquet JC (2010) Alternating proximal algorithm for
  blind image recovery.
\newblock In: Proceedings International Conference On Image Processing. Hong
  Kong, China, 26--29

\bibitem[{Bourget et~al.(2013)Bourget, Dauxois, Joubaud, and
  Odier}]{Bourget:JFM:13}
Bourget B, Dauxois T, Joubaud S, Odier P (2013) Experimental study of
  parametric subharmonic instability for internal plane waves.
\newblock Journal of Fluid Mechanics 723:1--20

\bibitem[{Brouzet et~al.(2017)Brouzet, Ermanyuk, Joubaud, Pillet, and
  Dauxois}]{Brouzet17}
Brouzet C, Ermanyuk E, Joubaud S, Pillet G, Dauxois T (2017) {Internal wave
  attractors: different scenarios of instability}.
\newblock Journal of Fluid Mechanics 811:544--568

\bibitem[{Buhler and Muller(2007)}]{Buhler2007}
Buhler O, Muller C (2007) Instability and focusing of internal tides in the
  deep ocean.
\newblock Journal of Fluid Mechanics 588:1--28

\bibitem[{Butman et~al.(2006)Butman, Alexander, Scotti, Beardsley, and
  Anderson}]{Butman:CSR:06}
Butman B, Alexander P, Scotti A, Beardsley R, Anderson S (2006) Large internal
  waves in massachusetts bay transport sediments offshore.
\newblock Continental Shelf Research 26(17):2029 -- 2049

\bibitem[{Cacchione et~al.(2002)Cacchione, Pratson, and Ogston}]{Cacchione2002}
Cacchione D, Pratson L, Ogston A (2002) The shaping of continental slopes by
  internal tides.
\newblock Science 296:724--727

\bibitem[{Cacchione and Wunsch({1974})}]{Cacchione:JFM:74}
Cacchione D, Wunsch C ({1974}) Experimental study of internal waves over a
  slope.
\newblock Journal of Fluid Mechanics {66}:223--239

\bibitem[{Chalamalla et~al.(2013)Chalamalla, Gayen, Scotti, and
  Sarkar}]{Chalamalla2013}
Chalamalla V, Gayen B, Scotti A, Sarkar S (2013) Turbulence during the
  reflection of internalgravity waves at critical and near-critical slopes.
\newblock Journal of Fluid Mechanics 729:47--68

\bibitem[{Daubechies et~al.(2011)Daubechies, Lu, and Wu}]{Daubechies:2011}
Daubechies I, Lu J, Wu HT (2011) Synchrosqueezed wavelet transforms: An
  empirical mode decomposition-like tool.
\newblock Applied and Computational Harmonic Analysis 30(2):243 -- 261

\bibitem[{Dauxois et~al.({2004})Dauxois, Didier, and Falcon}]{Dauxois:PoF:04}
Dauxois T, Didier A, Falcon E ({2004}) {Observation of near-critical reflection
  of internal waves in a stably stratified fluid}.
\newblock {Physics of Fluids} {16}({6}):{1936--1941}

\bibitem[{Dauxois et~al.(2018)Dauxois, Joubaud, Odier, and
  Venaille}]{Dauxois17}
Dauxois T, Joubaud S, Odier P, Venaille A (2018) {Instabilities of Internal
  Gravity Wave Beams}.
\newblock Annual Review of Fluid Mechanics 50:1--28

\bibitem[{Dauxois and Young(1999)}]{Dauxois:JFM:99}
Dauxois T, Young W (1999) {Near-critical reflection of internal waves}.
\newblock Journal of Fluid Mechanics {390}:271--295

\bibitem[{DeSilva et~al.({1997})DeSilva, Imberger, and Ivey}]{DeSilva:JFM:97}
DeSilva I, Imberger J, Ivey G ({1997}) {Localized mixing due to a breaking
  internal wave ray at a sloping bed}.
\newblock Journal of Fluid Mechanics {350}:1--27

\bibitem[{{Dragomiretskiy} and {Zosso}(2014)}]{Dragomiretskiy2014}
{Dragomiretskiy} K, {Zosso} D (2014) Variational mode decomposition.
\newblock IEEE Transactions on Signal Processing 62(3):531--544

\bibitem[{Foare et~al.(2019)Foare, Pustelnik, and Condat}]{Foare2019}
Foare M, Pustelnik N, Condat L (2019) Semi-linearized proximal alternating
  minimization for a discrete mumford-shah model.
\newblock In: IEEE Trans. on Image Processing. volume~29, 2176--2189

\bibitem[{Fortuin(1960)}]{Fortuin:JPS:60}
Fortuin JMH (1960) Theory and application of two supplementary methods of
  constructing density gradient columns.
\newblock Journal of Polymer Science 44(144):505--515

\bibitem[{Gayen and Sarkar(2010)}]{Gayen2010}
Gayen B, Sarkar S (2010) Turbulence during the generation of internal tide on a
  critical slope.
\newblock Physical Review Letters 104:218502

\bibitem[{Gostiaux et~al.({2006})Gostiaux, Dauxois, Didelle, Sommeria, and
  Viboud}]{Gostiaux:PoF:06}
Gostiaux L, Dauxois T, Didelle H, Sommeria J, Viboud S ({2006}) {Quantitative
  laboratory observations of internal wave reflection on ascending slopes}.
\newblock Physics of Fluids {18}({5}):056602

\bibitem[{Gostiaux et~al.({2007})Gostiaux, Didelle, Mercier, and
  Dauxois}]{Gostiaux:EF:07}
Gostiaux L, Didelle H, Mercier S, Dauxois T ({2007}) {A novel internal waves
  generator}.
\newblock Experiments in Fluids {42}({1}):123--130

\bibitem[{Hosegood et~al.(2004)Hosegood, Bonnin, and van
  Haren}]{Hosegood:GRL:04}
Hosegood P, Bonnin J, van Haren H (2004) Solibore-induced sediment resuspension
  in the faeroe-shetland channel.
\newblock Geophysical Research Letters 31(9)

\bibitem[{Huang et~al.(1998)Huang, Shen, Long, Wu, Shih, Zheng, Yen, Tung, and
  Liu}]{Huang:1998}
Huang NE, Shen Z, Long SR, Wu MC, Shih H, Zheng Q, Yen NC, Tung CC, Liu HH
  (1998) The {E}mpirical {M}ode {D}ecomposition and the {H}ilbert spectrum for
  nonlinear and nonstationary time series analysis.
\newblock Proceedings of the Royal Society 454:903--995

\bibitem[{Ivey et~al.({2008})Ivey, Winters, and Koseff}]{Ivey:ARFM:2008}
Ivey GN, Winters KB, Koseff JR ({2008}) {Density stratification, turbulence,
  but how much mixing?}
\newblock {Annual Review of Fluid Mechanics} {40}:169--184

\bibitem[{Kataoka and Akylas(2020)}]{Kataoka:2020}
Kataoka T, Akylas T (2020) Viscous reflection of internal waves from a slope.
\newblock Physical Review Fluids 5:014803

\bibitem[{Kistovich and Chashechkin({1995})}]{Kistovich:JAMM:95}
Kistovich Y, Chashechkin Y ({1995}) {The reflection of beams of internal
  gravity waves at a flat rigid surface}.
\newblock Journal of Applied Mathematics and Mechanics {59}({4}):{579--585}

\bibitem[{Mercier et~al.({2008})Mercier, Garnier, and Dauxois}]{Mercier:PoF:08}
Mercier MJ, Garnier NB, Dauxois T ({2008}) {Reflection and diffraction of
  internal waves analyzed with the Hilbert transform}.
\newblock {Physics of Fluids} {20}({8}):086601

\bibitem[{Mercier et~al.({2010})Mercier, Martinand, Mathur, Gostiaux, Peacock,
  and Dauxois}]{Mercier:JFM:10}
Mercier MJ, Martinand D, Mathur M, Gostiaux L, Peacock T, Dauxois T ({2010})
  {New wave generation}.
\newblock J Fluid Mech {657}:308--334

\bibitem[{Oster and Yamamoto(1963)}]{Oster:CR:63}
Oster G, Yamamoto M (1963) Density gradient techniques.
\newblock Chemical Reviews 63(3):257--268

\bibitem[{Ouriemi et~al.(2007)Ouriemi, Aussillous, Medale, Peysson, and
  Guazzelli}]{Ouriemi2007}
Ouriemi M, Aussillous P, Medale M, Peysson Y, Guazzelli E (2007) Determination
  of the critical shields number for particle erosion in laminar flow.
\newblock Physics of Fluids 19(6):061706

\bibitem[{Peacock and Tabaei({2005})}]{Peacock:PoF:05}
Peacock T, Tabaei A ({2005}) {Visualization of nonlinear effects in reflecting
  internal wave beams}.
\newblock {Physics of Fluids} {17}({6})

\bibitem[{Phillips(1966)}]{Phillips:66}
Phillips OM (1966) The dynamics of the upper ocean.
\newblock Cambridge University Press

\bibitem[{Phillips({1970})}]{Phillips:DSR:70}
Phillips OM ({1970}) On flows induced by diffusion in stably stratified fluids.
\newblock Deep Sea Research {17}({3}):435--443

\bibitem[{Pustelnik et~al.(2014)Pustelnik, Borgnat, and
  Flandrin}]{Pustelnik:2014}
Pustelnik N, Borgnat P, Flandrin P (2014) Empirical mode decomposition
  revisited by multicomponent non smooth convex optimization.
\newblock Signal Processing 102:313--331

\bibitem[{Quaresma et~al.({2007})Quaresma, Vitorino, Oliveira, and
  da~Silva}]{Quaresma:MG:07}
Quaresma L, Vitorino J, Oliveira A, da~Silva JCB ({2007}) {Evidence of sediment
  resuspension by nonlinear internal waves on the western Portuguese
  mid-shelf}.
\newblock Marine Geology {246}({2-4}):{123--143}

\bibitem[{Rilling et~al.(2003)Rilling, Flandrin, and
  Gon{\c{c}}alv\`es}]{Rilling:2003}
Rilling G, Flandrin P, Gon{\c{c}}alv\`es P (2003) On {E}mpirical {M}ode
  {D}ecomposition and its algorithms.
\newblock In: IEEE-EURASIP Workshop on Nonlinear Signal and Image Processing
  ({NSIP}-03)

\bibitem[{Rodenborn et~al.(2011)Rodenborn, Kiefer, Zhang, and
  Swinney}]{Rodenborn:2011}
Rodenborn B, Kiefer H, Zhang H, Swinney H (2011) Harmonic generation by
  reflecting internal waves.
\newblock Physics of Fluids 23:026601

\bibitem[{Sarkar and Scotti(2017)}]{SarkarARFM2017}
Sarkar S, Scotti A (2017) From topographic internal gravity waves to
  turbulence.
\newblock Annual Review of Fluid Mechanics 49:195--220

\bibitem[{Schmitt et~al.(2015)Schmitt, Horne, Pustelnik, Joubaud, and
  Odier}]{Schmitt:EUSIPCO:15}
Schmitt J, Horne E, Pustelnik N, Joubaud S, Odier P (2015) An improved
  variational mode decomposition method for internal waves separation.
\newblock In: Signal Processing Conference (EUSIPCO), 2015 23rd European.
  1935--1939

\bibitem[{Scotti({2011})}]{Scotti:JFM:11}
Scotti A ({2011}) {Inviscid critical and near-critical reflection of internal
  waves in the time domain}.
\newblock Journal of Fluid Mechanics {674}:464--488

\bibitem[{Tabaei et~al.({2005})Tabaei, Akylas, and Lamb}]{Tabaei:JFM:05}
Tabaei A, Akylas T, Lamb K ({2005}) {Nonlinear effects in reflecting and
  colliding internal wave beams}.
\newblock Journal of Fluid Mechanics {526}:{217--243}

\bibitem[{Thorpe(1987)}]{Thorpe:JFM:1987}
Thorpe SA (1987) On the reflection of a strain of finite-amplitude internal
  waves from a uniform slope.
\newblock {Journal of Fluid Mechanics} 178:279--302.

\bibitem[{Wunsch({1969})}]{Wunsch:JFM:69}
Wunsch C ({1969}) Progressive internal waves on slopes.
\newblock J Fluid Mech {35}({1}):131--144

\bibitem[{Zhang et~al.({2008})Zhang, King, and Swinney}]{Zhang:PRL:08}
Zhang HP, King B, Swinney HL ({2008}) {Resonant generation of internal waves on
  a model continental slope}.
\newblock {Phys Rev Lett} {100}({24}):244504

\bibitem[{Zosso et~al.(2017)Zosso, Dragomiretskiy, Bertozzi, and
  Weiss}]{Zosso2017}
Zosso D, Dragomiretskiy K, Bertozzi A, Weiss P (2017) Two-dimensional compact
  variational mode decomposition spatially compact and spectrally sparse image
  decomposition and segmentation.
\newblock volume~58, 294–320

\end{thebibliography}

%
%

\end{document}